\documentclass[12pt]{article}
\pdfoutput=1
\usepackage{amsmath,amssymb,amsthm,amsxtra,overpic,bbm,bm,epsfig,subfigure}
\usepackage{hyperref}
\usepackage{graphicx}
\usepackage{color}
\usepackage{array}
\newcommand{\PreserveBackslash}[1]{\let\temp=\\#1\let\\=\temp}
\newcolumntype{C}[1]{>{\PreserveBackslash\centering}p{#1}}
\newcolumntype{R}[1]{>{\PreserveBackslash\raggedleft}p{#1}}
\newcolumntype{L}[1]{>{\PreserveBackslash\raggedright}p{#1}}
\usepackage{comment}
\usepackage{epstopdf}
\usepackage{float}
\usepackage{multirow}
\numberwithin{equation}{section}
\usepackage{cite}
\usepackage{hyperref}
\usepackage{slashed,stmaryrd}
\usepackage{bbm}
\usepackage{lscape}%
\usepackage{array}
\usepackage{booktabs}%

\def\thefootnote{\fnsymbol{footnote}}

\begin{document}

\vspace{0.2cm}

\begin{center}
{\Large\bf Double Covering of the Modular $A^{}_5$ Group and Lepton Flavor Mixing in the Minimal Seesaw Model}
\end{center}

\vspace{0.2cm}

\begin{center}
{\bf Xin Wang}~$^{a,~b}$~\footnote{E-mail: wangx@ihep.ac.cn},
\quad
{\bf Bingrong Yu}~$^{a,~b}$~\footnote{E-mail: yubr@ihep.ac.cn},
\quad
{\bf Shun Zhou}~$^{a,~b}$~\footnote{E-mail: zhoush@ihep.ac.cn (corresponding author)}
\\
\vspace{0.2cm}
{\small
$^a$Institute of High Energy Physics, Chinese Academy of Sciences, Beijing 100049, China\\
$^b$School of Physical Sciences, University of Chinese Academy of Sciences, Beijing 100049, China}
\end{center}

\vspace{1.5cm}

\begin{abstract}
In this paper, we investigate the double covering of modular $\Gamma^{}_5 \simeq A^{}_5$ group and derive all the modular forms of weight one for the first time. The modular forms of higher weights are also explicitly given by decomposing the direct products of weight-one forms. For the double covering group $\Gamma^\prime_5 \simeq A^\prime_5$, there exist two inequivalent two-dimensional irreducible representations, into which we can assign two right-handed neutrino singlets in the minimal seesaw model. Two concrete models with such a salient feature have been constructed to successfully explain lepton mass spectra and flavor mixing pattern. The allowed parameter space for these two minimal scenarios has been numerically explored, and analytically studied with some reasonable assumptions.
\end{abstract}


\def\thefootnote{\arabic{footnote}}
\setcounter{footnote}{0}

\newpage

\section{Introduction}\label{sec:intro}

The discovery of neutrino oscillations calls for new physics beyond the standard model (SM) to dynamically generate tiny neutrino masses and significant lepton flavor mixing~\cite{Xing:2011zza, Xing:2019vks, Esteban:2020cvm}. In order to account for tiny neutrino masses, one can extend the SM with three right-handed neutrino singlets $N^{}_{i{\rm R}}$ (for $i = 1, 2, 3$) such that the gauge-invariant Lagrangian for lepton masses and flavor mixing is given by
\begin{eqnarray}
-{\cal L}^{}_{\rm lepton} = \overline{\ell^{}_{\rm L}} Y^{}_l H E^{}_{\rm R} + \overline{\ell^{}_{\rm L}} Y^{}_\nu \widetilde{H} N^{}_{\rm R} + \frac{1}{2} \overline{N^{\rm C}_{\rm R}} M^{}_{\rm R} N^{}_{\rm R} + {\rm h.c.} \; ,
\label{eq:lagseesaw}
\end{eqnarray}
where $\ell^{}_{\rm L} \equiv (\nu^{}_{\rm L}, E^{}_{\rm L})^{\rm T}$ and $H \equiv (H^+, H^0)^{\rm T}$ stand respectively for the left-handed lepton doublet and the Higgs doublet, $\widetilde{H} \equiv {\rm i}\sigma^{}_2 H^*$ and $N^{\rm C}_{\rm R} \equiv {\cal C}\overline{N^{}_{\rm R}}^{\rm T}$ with ${\cal C} \equiv {\rm i}\gamma^2 \gamma^0$ being the charge-conjugation matrix have been defined, $Y^{}_l$ and $Y^{}_\nu$ are the charged-lepton and Dirac neutrino Yukawa coupling matrices, and $M^{}_{\rm R}$ is the Majorana mass matrix of right-handed neutrinos. After the spontaneous symmetry breaking, the charged-lepton and Dirac neutrino mass matrices are given by $M^{}_l \equiv Y^{}_l v/{\sqrt{2}}$ and $M^{}_{\rm D} \equiv Y^{}_\nu v/\sqrt{2}$, respectively, with $v \approx 246~{\rm GeV}$ being the vacuum expectation value (vev) of the SM Higgs field. Given ${\cal O}(M^{}_{\rm D}) \approx 100~{\rm GeV}$ and ${\cal O}(M^{}_{\rm R}) \approx 10^{14}~{\rm GeV}$, the effective Majorana mass matrix of three light neutrinos is determined by the seesaw formula $M^{}_\nu \approx - M^{}_{\rm D} M^{-1}_{\rm R} M^{\rm T}_{\rm D}$~\cite{Minkowski, Yanagida, Gell-Mann, Glashow, Mohapatra} and turns out to be on the right order, i.e., ${\cal O}(M^{}_\nu) \approx 0.1~{\rm eV}$.

Although the smallness of light Majorana neutrino masses can naturally be ascribed to the heaviness of right-handed neutrino singlets in the canonical seesaw model~\cite{Minkowski, Yanagida, Gell-Mann, Glashow, Mohapatra}, the observed pattern of lepton flavor mixing remains completely unexplained~\cite{Xing:2019vks}. According to the latest global-fit analysis of all neutrino oscillation data~\cite{Esteban:2020cvm}, from which the current knowledge on neutrino mass-squared differences $\Delta m^2_{ij} \equiv m^2_i - m^2_j$ (for $ij = 21, 31, 32$), three mixing angles $\{\theta^{}_{12}, \theta^{}_{13}, \theta^{}_{23}\}$ and the Dirac-type CP-violating phase $\delta$ has been summarized in Table~\ref{table:gfit}, we can observe that the flavor mixing angles $\theta^{}_{12} \approx 33^\circ$, $\theta^{}_{13} \approx 8.6^\circ$ and $\theta^{}_{23} \approx 49^\circ$ in the lepton sector are dramatically different from those in the quark sector~\cite{PDG2020}. An attractive way to understand the lepton flavor mixing is to impose discrete flavor symmetries on the seesaw model that is further extended with a number of scalar fields, which may transform nontrivially under the flavor symmetry group. See, e.g., Refs.~\cite{Altarelli:2010gt, Ishimori:2010au, King:2013eh, King:2014nza}, for recent reviews on discrete flavor symmetries and their applications to lepton flavor mixing and CP violation.

Apart from the discrete flavor symmetries, the finite modular symmetries have been recently implemented to account for lepton flavor mixing~\cite{Altarelli:2005yx, deAdelhartToorop:2011re, Feruglio:2017spp}. In the frameworks of string theories and supersymmetric field theories, the modular invariance and its connection to discrete flavor symmetries have been excellently elaborated in Ref.~\cite{Feruglio:2017spp}. In addition, the practical applications of finite modular groups $\Gamma^{}_N$ (for $N = 2,3,4,5,\cdots$) to the model building of neutrino masses and flavor mixing can be found in the vast literature\cite{Kobayashi:2018vbk, Kobayashi:2018wkl, Kobayashi:2019rzp, Okada:2019xqk, Kobayashi:2018scp, Criado:2018thu, deAnda:2018ecu, Okada:2018yrn, Nomura:2019jxj, Nomura:2019yft, Ding:2019zxk, Nomura:2019lnr, Okada:2019mjf, Asaka:2019vev, Zhang:2019ngf, Penedo:2018nmg, Novichkov:2018ovf, Okada:2019lzv, Novichkov:2018nkm, Ding:2019xna, Criado:2019tzk, Novichkov:2019sqv, deMedeirosVarzielas:2019cyj, King:2019vhv, Kobayashi:2019mna, Kobayashi:2019xvz, Novichkov:2018yse, Gui-JunDing:2019wap, Okada:2019uoy, Baur:2019kwi, Baur:2019iai,Wang:2019ovr,Kobayashi:2019uyt,Nomura:2019xsb,Kobayashi:2019gtp,Wang:2019xbo,Nilles:2020nnc,Kobayashi:2020hoc,King:2020qaj,Abe:2020vmv,Ohki:2020bpo,Okada:2020dmb,Ding:2020msi,Okada:2020rjb,Kikuchi:2020frp,Nilles:2020tdp,Behera:2020sfe,Nomura:2020opk,Wang:2020dbp,Liu:2020msy,Nomura:2020cog,deMedeirosVarzielas:2020kji,Baur:2020jwc,Aoki:2020eqf,Asaka:2020tmo,Okada:2020ukr,Nagao:2020snm,Ding:2020zxw}. Meanwhile, the double covering of $\Gamma^{}_N$ has also been discussed in~\cite{Liu:2019khw,Lu:2019vgm,Novichkov:2020eep,Liu:2020akv,Kikuchi:2020nxn}. In the present paper, we investigate the double covering of the modular $A^{}_5$ symmetry group, i.e., $\Gamma^\prime_5 \simeq A^\prime_5$, which has not yet been explored in the previous works. For the finite modular group $\Gamma^\prime_5$, the modular forms with both even and odd weights are present. The explicit expressions of the modular forms of weights up to six in the nontrivial representations of $\Gamma^\prime_5$ are derived for the first time. Interestingly, there are two-dimensional irreducible representations $\widehat{\bf 2}$ and $\widehat{\bf 2}^\prime$ for the double covering group $\Gamma^\prime_5 \simeq A^\prime_5$, which are however absent for the $\Gamma^{}_5 \simeq A^{}_5$ group. Motivated by this observation, we further apply the $\Gamma^\prime_5$ group to the minimal seesaw model (MSM)~\cite{Kleppe:1995zz, Ma:1998zg, King:1999mb, King:2002nf, Frampton:2002qc, Guo:2006qa, Xing:2020ald} and assign two right-handed neutrino singlets into the two-dimensional representation of $\Gamma^\prime_5$. Two concrete examples in the MSM have been given to explain the observed pattern of lepton flavor mixing.

The remaining part of this paper is structured as follows. In Sec.~\ref{sec:double}, the double covering group $\Gamma^\prime_5 \simeq A^\prime_5$ of the modular $\Gamma^{}_5 \simeq A^{}_5$ group is examined and the modular forms of weights up to six are explicitly given. The applications of the $\Gamma^\prime_5$ group to the MSM are explored in Sec.~\ref{sec:msm} and two concrete models are built to account for lepton flavor mixing as observed in neutrino oscillation experiments. The exact numerical results of the allowed parameter space are presented, while the approximate analytical results are also derived in order to understand the flavor structures of the charged-lepton and neutrino mass matrices. In Sec.~\ref{sec:sum}, we summarize our main results. Finally, the basic properties of the finite group $A^\prime_5$ are presented in Appendices~\ref{app:A} and \ref{app:B}.

\begin{table}[t]
	\begin{center}
		\vspace{-0.25cm} \caption{The best-fit values, the
			1$\sigma$ and 3$\sigma$ intervals, together with the values of $\sigma^{}_{i}$ being the symmetrized $1\sigma$ uncertainties, for three neutrino mixing angles $\{\theta^{}_{12}, \theta^{}_{13}, \theta^{}_{23}\}$, two neutrino mass-squared differences $\{\Delta m^2_{21}, \Delta m^2_{31}~{\rm or}~\Delta m^2_{32}\}$ and the Dirac CP-violating phase $\delta$ from a global-fit analysis of current experimental data~\cite{Esteban:2020cvm}.} \vspace{0.5cm}
		\begin{tabular}{c|c|c|c|c}
			\hline
			\hline
			Parameter & Best fit & 1$\sigma$ range &  3$\sigma$ range & $\sigma^{}_{i}$ \\
			\hline
			\multicolumn{5}{c}{Normal neutrino mass ordering
				$(m^{}_1 < m^{}_2 < m^{}_3)$} \\ \hline
			$\sin^2_{}\theta^{}_{12}$
			& $0.304$ & 0.292 --- 0.317 &  0.269 --- 0.343 & 0.0125 \\
			$\sin^2_{}\theta^{}_{13}$
			& $0.02221$ & 0.02159 --- 0.02289 &  0.02034 --- 0.02430  & 0.00065 \\
			$\sin^2_{}\theta^{}_{23}$
			& $0.570$  & 0.546 --- 0.588 &  0.407 --- 0.618  & 0.021 \\
			$\delta/{}^\circ_{}$ &  $195$ & 170 --- 246 &  107 --- 403 & 38 \\
			$\Delta m^2_{21}/ (10^{-5}~{\rm eV}^2)$ &  $7.42$ & 7.22 --- 7.63 & 6.82 --- 8.04 & 0.205 \\
			$\Delta m^2_{31}/(10^{-3}~{\rm eV}^2)$ &  $+2.514$ & +2.487 --- +2.542 & +2.431 --- +2.598 & 0.0275 \\\hline
			\multicolumn{5}{c}{Inverted neutrino mass ordering
				$(m^{}_3 < m^{}_1 < m^{}_2)$} \\ \hline
			$\sin^2_{}\theta^{}_{12}$
			& $0.304$ & 0.292 --- 0.317 &  0.269 --- 0.343 & 0.0125\\
			$\sin^2_{}\theta^{}_{13}$
			& $0.02240$ & 0.02178 --- 0.02302 &  0.02053 --- 0.02436 & 0.00062 \\
			$\sin^2_{}\theta^{}_{23}$
			& $0.575$  & 0.554 --- 0.592 &  0.411 --- 0.621 & 0.019  \\
			$\delta/{}^\circ_{}$ &  $286$ & 254 --- 313 &  192 --- 360 & 29.5 \\
			$\Delta m^2_{21}/ (10^{-5}~{\rm eV}^2)$ &  $7.42$ & 7.22 --- 7.63 & 6.82 --- 8.04 & 0.205 \\
			$\Delta m^2_{32}/(10^{-3}~{\rm eV}^2)$ &  $-2.497$ & $-2.525$ --- $-2.469$  & $-2.583$ --- $-2.412$ & $0.028$ \\ \hline\hline
		\end{tabular}
		\label{table:gfit}
	\end{center}
\end{table}

\section{Double Covering Group}\label{sec:double}

\subsection{Double covering of the modular group}

In this subsection, we introduce the double covering of the modular group and explain why the modular group $\Gamma \simeq {\rm SL}(2,\mathbb{Z})$ and its principal congruence subgroups $\Gamma(N)$ (for $N >2$ being positive integers) can accommodate the modular forms with odd weights. Although the basic properties of the modular group can be found in the existing literature~\cite{Feruglio:2017spp} and mathematical monographs~\cite{Lang, Miyake, Diamond, Schultz:2015}, a concise introduction will be helpful in establishing our notations and conventions for the subsequent discussions. First, let us recall the definition of the modular group $\Gamma \simeq {\rm SL}(2,\mathbb{Z})$, namely,
\begin{eqnarray}
\Gamma \equiv \left\{\left(\begin{matrix}
a && b \\
c && d
\end{matrix}\right) \bigg| a, b, c, d \in \mathbb{Z}\; ,~ a d - b c = 1 \right\} \; ,
\label{eq:modgr}
\end{eqnarray}
which is generated by $S$, $T$ and $R$ satisfying $S^2_{} = R$, $(ST)^3_{} = \mathbb{I}$, $R^2_{} = \mathbb{I}$ and $RT =T R$. More explicitly, the matrix representations of these three generators are given by
\begin{eqnarray}
S = \left(\begin{matrix}
0 && 1 \\
-1 && 0
\end{matrix}\right) \; , \quad T = \left(\begin{matrix}
1 && 1 \\
0 && 1
\end{matrix}\right) \; , \quad R = \left(\begin{matrix}
-1 && 0 \\
0 && -1
\end{matrix}\right) \; .
\label{eq:STmatrix}
\end{eqnarray}
Suppose that $\gamma$ is an element of the modular group $\Gamma$. Then the modular transformations of the complex parameter $\tau$ in the upper half of the complex plane $\mathcal{H}=\{\tau \in \mathbb{C} \mid \operatorname{Im} \tau>0\}$ and the chiral supermultiplet $\chi^{(I)}_{}$ can be defined as
\begin{eqnarray}
\gamma: \tau \rightarrow \dfrac{a \tau + b}{c \tau + d} \; , \quad
\chi^{(I)}_{} \rightarrow (c \tau +d )^{-k^{}_I} \rho^{(I)}_{} (\gamma) \chi^{(I)} _{} \; ,
\label{eq:lintra}
\end{eqnarray}
where $k^{}_I$ is the weight of the chiral supermultiplet and $\rho^{(I)}_{}(\gamma)$ denotes the representation matrix of $\gamma$. Obviously, the transformations of $\tau$ induced by $\gamma$ and $-\gamma$ are actually identical, i.e., the modulus $\tau$ does not transform under $R$. Hence one can define the so-called inhomogeneous modular group as $\overline{\Gamma} \simeq {\rm PSL}(2, \mathbb{Z}) \equiv {\rm SL}(2,\mathbb{Z})/\{\mathbb{I}, -\mathbb{I}\}$ with $\mathbb{I}$ being the identity element, for which two generators are found to be $S:~\tau \rightarrow -1/\tau$ and $T:~\tau \rightarrow \tau+1$. On the other hand, the modular group has an important class of subgroups, i.e., the principal congruence subgroups, whose exact definition is as follows
\begin{eqnarray}
\Gamma(N)=\left\{\left(\begin{matrix}
a && b \\
c && d
\end{matrix}\right) \in {\rm SL}(2, \mathbb{Z})\; , \quad\left(\begin{matrix}
a && b \\
c && d
\end{matrix}\right)=\left(\begin{matrix}
1 && 0 \\
0 && 1
\end{matrix}\right) \; (\bmod\; N)\right\} \; ,
\label{eq:prinsubg}
\end{eqnarray}
for a given positive integer $N$. In a similar way, one can introduce $\overline{\Gamma}(N) \equiv \Gamma(N)/\{\mathbb{I}, -\mathbb{I}\}$, and the quotient group $\Gamma^{}_N \equiv \overline{\Gamma}/\overline{\Gamma}(N)$, which are generated by $S$ and $T$ satisfying the identities $S^{2}_{} = (S T)^{3}_{} = T^{N}_{} = \mathbb{I}$. However, as has been pointed out in \cite{Nilles:2020nnc,Novichkov:2020eep}, matter fields in modular-invariant theories are generally allowed to transform under $R$. Therefore, we should consider $\Gamma$ rather than $\overline{\Gamma}$ as the symmetry group in such theories. As a consequence, the finite modular group $\Gamma^{}_N$ will be extended to its double covering group, defined as $\Gamma^\prime_N \equiv \Gamma/\Gamma(N)$, which is generated by $S$, $T$ and $R$ that obey the following identities
\begin{eqnarray}
S^2_{}=R \; , \quad (ST)^3_{} = \mathbb{I} \; , \quad T^N_{} = \mathbb{I} \; , \quad R^2_{} = \mathbb{I} \; , \quad RT =TR \; .
\label{eq:genedou}
\end{eqnarray}

Then, we consider the modular forms. By definition, the modular form $f(\tau)$ of level $N$ and weight $k$ is a holomorphic function of $\tau$, and it transforms under $\Gamma(N)$ as
\begin{eqnarray}
f\left(\gamma\tau\right)=(c \tau+d)^{k} f(\tau) \; , \quad \gamma \in \Gamma(N) \; ,
\label{eq:modform}
\end{eqnarray}
where $k \geq 0$ is an integer. Note that $-\mathbb{I}$ belongs to $\Gamma(N)$ for $N=1,2$, leading to $f(\tau) = (-1)^k_{}f(\tau)$ if we substitute $\gamma = -\mathbb{I}$ into Eq.~(\ref{eq:modform}). Therefore, the nonzero modular forms with odd weights can only exist in $\Gamma(N)$ with $N>2$. As has been proved in Ref.~\cite{Liu:2019khw}, for a given modular space ${\cal M}^{}_k\left[\Gamma(N)\right]$, the modular forms can always be decomposed into several multiplets that transform as irreducible unitary representations of the finite modular group $\Gamma^{\prime}_{N}$. To be more precise, we can always find a proper basis for the modular space ${\cal M}^{}_k\left[\Gamma(N)\right]$ such that a modular multiplet $Y^{(k)}_{\bf r} = (f^{}_1(\tau),f^{}_2(\tau), \cdots)^{\rm T}_{}$ in the representation ${\bf r}$ satisfies the following equation
\begin{eqnarray}
Y^{(k)}_{\bf r}(\gamma\tau) = (c\tau+d)^k_{}\rho^{}_{\bf r}(\gamma)Y^{(k)}_{\bf r}(\tau)\;, \quad \gamma \in \Gamma\; ,
\label{eq:modformtran}
\end{eqnarray}
where $\rho^{}_{\bf r}(\gamma)$ denotes the representation matrix of $\gamma$, and $\rho^{}_{\bf r}(\gamma)=1$ for $\gamma \in \Gamma(N)$. In particular, for the generators $\gamma = S$ and $T$, we get
\begin{eqnarray}
Y^{(k)}_{\mathbf{r}}(S \tau)=(-\tau)^{k}_{} \rho_{\mathbf{r}}(S) Y^{(k)}_{\mathbf{r}}(\tau) \; , \quad Y^{(k)}_{\mathbf{r}}(T \tau)=\rho_{\mathbf{r}}(T) Y^{(k)}_{\mathbf{r}}(\tau) \; .
\label{eq:ST}
\end{eqnarray}
Now that two elements $\gamma$ and $S^2_{}\gamma$ correspond to the same fractional linear transformation of $\tau$ in Eq.~(\ref{eq:lintra}), we can substitute them into Eq.~(\ref{eq:modformtran}) and arrive at
\begin{eqnarray}
Y^{(k)}_{\mathbf{r}}(\gamma \tau)&=&(c \tau+d)^{k}_{} \rho^{}_{\mathbf{r}}(\gamma) Y^{(k)}_{\mathbf{r}}(\tau) \; , \nonumber \\
Y^{(k)}_{\mathbf{r}}\left(S^{2}_{} \gamma \tau\right)&=&(-1)^{k}_{}(c \tau+d)^{k}_{} \rho^{}_{\mathbf{r}}\left(S^{2}_{} \gamma\right) Y^{(k)}_{\mathbf{r}}(\tau) \; ,
\label{eq:gaSga}
\end{eqnarray}
which should be identical, leading to the relation
\begin{eqnarray}
\rho^{}_{\bf r}(S^{2}_{}\gamma)=(-1)^k_{}\rho^{}_{\bf r}(\gamma) \; .
\label{eq:gaSgae}
\end{eqnarray}
Since $S^2_{} = \mathbb{I}$ in the finite modular group $\Gamma^{}_{N}$, Eq.~(\ref{eq:gaSgae}) implies $\rho^{}_{\bf r}(\gamma) = (-1)^k_{} \rho^{}_{\bf r}(\gamma)$, which is expected to hold only for $k$ being an even integer. However, in the double covering group $\Gamma^\prime_{N}$, we can see that the generator $R = S^2_{}$ fulfills the relations $\rho^{}_{\mathbf{r}}(R) = \rho^{}_{\mathbf{r}}(\mathbb{I}) = \mathbb{I}$ for an even $k$ and $\rho^{}_{\mathbf{r}}(R) = -\rho^{}_{\mathbf{r}}(\mathbb{I}) = - \mathbb{I}$ for an odd $k$. Therefore, we demonstrate that the modular group $\Gamma$ and its principal congruence subgroup $\Gamma(N)$ with $N>2$ have to be considered for the modular forms of both even and odd weights.

\subsection{The group $\Gamma^\prime_5 \simeq A^\prime_5$}

In this paper, we focus on the finite modular group $\Gamma^{}_5 \simeq A^{}_5$ and its double covering group $\Gamma^\prime_5 \simeq A^\prime_5$, corresponding to the specific case of $N = 5$. The basic properties of the finite group $A^\prime_5$ have already been studied in the existing literature~\cite{Shirai:1992,Everett:2010rd,Hashimoto:2011tn,Chen:2011dn}, so we just briefly summarize the key points relevant for our later discussions. The $A^{\prime}_{5}$ group has 120 elements, which can be produced by three generators $S$, $T$ and $R$ satisfying the identities in Eq.~(\ref{eq:genedou}) for $N = 5$.
All the 120 elements can be divided into nine conjugacy classes, indicating that $A^{\prime}_{5}$ has nine distinct irreducible representations, which are normally denoted as ${\bf 1}$, $\widehat{\bf 2}$, $\widehat{\bf 2}^{\prime}_{}$, ${\bf 3}$, ${\bf 3}^{\prime}_{}$, ${\bf 4}$, $\widehat{\bf 4}$, ${\bf 5}$ and $\widehat{\bf 6}$ by their dimensions. The conjugacy classes and character table of $A^\prime_5$, together with the representation matrices of all three generators $S$, $T$ and $R$ in the irreducible representations, are explicitly given in Appendix~\ref{app:A}. Notice that the representations ${\bf 1}$, ${\bf 3}$, ${\bf 3}^{\prime}_{}$, ${\bf 4}$ and ${\bf 5}$ with $R = \mathbb{I}$ coincide with those for $A^{}_{5}$, whereas $\widehat{\bf 2}$, $\widehat{\bf 2}^{\prime}_{}$, $\widehat{\bf 4}$ and $\widehat{\bf 6}$ are unique for $A^\prime_5$ with $R = -\mathbb{I}$. In addition, the decomposition rules of the Kronecker products of any two nontrivial irreducible representations can be found in Appendix~\ref{app:B}, and will be frequently used in the subsequent discussions.

\subsection{The modular space of $\Gamma(5)$}

To construct the modular forms that transform nontrivially under $\Gamma^{\prime}_{5}$, which is isomorphic to $A^\prime_5$, we need first to find out the modular space of $\Gamma(5)$. For a given non-negative integer $k$, the modular space ${\cal M}^{}_{k}\left[\Gamma(5)\right]$ of weight $k$ for $\Gamma(5)$ contains $5k+1$ linearly independent modular forms, which can be regarded as the basis vectors of the modular space. According to Ref.~\cite{Schultz:2015}, we have
\begin{eqnarray}
{\cal M}^{}_k\left[\Gamma(5)\right] = \bigoplus^{}_{\substack{a+b = 5k \\a, b \geq 0}} \mathbb{C}\, \frac{\eta(5 \tau)^{15 k}}{\eta(\tau)^{3 k}} \, {\mathfrak k}^a_{\frac{1}{5},\frac{0}{5}}(5\tau) \, {\mathfrak k}^b_{\frac{2}{5},\frac{0}{5}}(5\tau) \; ,
\label{eq:G5basis}
\end{eqnarray}
where $\eta(\tau)$ is the Dedekind eta function
\begin{eqnarray}
\eta(\tau)=q^{1 / 24}_{} \prod_{n=1}^{\infty}\left(1-q^{n}_{}\right) \; ,
\label{eq:Dedekindeta}
\end{eqnarray}
with $q \equiv e^{2 {\rm i} \pi \tau}_{}$, and ${\mathfrak k}^{}_{r^{}_1,r^{}_2}(\tau)$ is the Klein form
\begin{eqnarray}
\mathfrak{k}^{}_{r^{}_{1}, r^{}_{2}}(\tau)= q_{z}^{(r^{}_{1}-1) / 2}\left(1-q^{}_{z}\right) \times \prod_{n=1}^{\infty}\left(1-q^{n}_{} q^{}_{z}\right)\left(1-q^{n}_{} q_{z}^{-1}\right)\left(1-q^{n}_{}\right)^{-2}_{} \; ,
\label{eq:Kexpansion}
\end{eqnarray}
with $(r^{}_1,r^{}_2)$ being a pair of rational numbers in the domain of ${\mathbb Q}^2_{}-{\mathbb Z}^2_{}$, $z \equiv \tau r^{}_{1} + r^{}_{2}$ and $q^{}_z \equiv e^{2 {\rm i} \pi z}$. Under the transformations of $S$ and $T$, the eta function and the Klein form change as follows
\begin{eqnarray}
\begin{array}{cclcl}
S &: ~\quad & \eta(\tau) \rightarrow \sqrt{-{\rm i} \tau} \eta(\tau) \; , &~\quad &\mathfrak{k}^{}_{r^{}_{1}, r^{}_{2}}(\tau) \rightarrow -\dfrac{1}{\tau}\, \mathfrak{k}^{}_{-r^{}_{2}, r^{}_{1}}(\tau) \; , \\
T &: ~\quad &\eta(\tau) \rightarrow e^{\rm{i} \pi / 12}_{} \eta(\tau) \; , & ~\quad &\mathfrak{k}^{}_{r^{}_{1}, r^{}_{2}}(\tau) \rightarrow \mathfrak{k}^{}_{r^{}_{1}, r^{}_{1}+r^{}_{2}}(\tau) \; .
\end{array}
\label{eq:etaklein}
\end{eqnarray}
More information about the properties of the Klein form ${\mathfrak k}^{}_{r^{}_1, r^{}_2}(\tau)$ can be found in Refs.~\cite{Schultz:2015, Ding:2019xna}, and thus will not be further discussed here.

Then we take $k = 1$ and find out the modular forms of the lowest weight. With the help of Eq.~(\ref{eq:G5basis}), it is straightforward to obtain the basis vectors of the modular space ${\cal M}^{}_1 \left[\Gamma(5)\right]$, i.e.,
\begin{eqnarray}
\begin{array}{lcl}
\widehat{e}^{}_{1}(\tau) = \dfrac{\eta^{15}(5 \tau)}{\eta^{3}(\tau)} \, \mathfrak{k}^{5}_{\frac{2}{5}, \frac{0}{5}}(5 \tau) \; , & ~\quad & \widehat{e}^{}_{2}(\tau) = \dfrac{\eta^{15}(5 \tau)}{\eta^{3}(\tau)} \, \mathfrak{k}^{}_{\frac{1}{5}, \frac{0}{5}}(5 \tau) \,  \mathfrak{k}^{4}_{\frac{2}{5}, \frac{0}{5}}(5 \tau) \; , \\
\widehat{e}^{}_{3}(\tau) = \dfrac{\eta^{15}(5 \tau)}{\eta^{3}(\tau)} \, \mathfrak{k}^{2}_{\frac{1}{5}, \frac{0}{5}}(5 \tau) \, \mathfrak{k}^{3}_{\frac{2}{5}, \frac{0}{5}}(5 \tau) \; , & ~\quad & \widehat{e}^{}_{4}(\tau) = \dfrac{\eta^{15}(5 \tau)}{\eta^{3}(\tau)} \, \mathfrak{k}^{3}_{\frac{1}{5}, \frac{0}{5}}(5 \tau) \, \mathfrak{k}^{2}_{\frac{2}{5}, \frac{0}{5}}(5 \tau) \; , \\
\widehat{e}^{}_{5}(\tau) = \dfrac{\eta^{15}(5 \tau)}{\eta^{3}(\tau)} \, \mathfrak{k}^{4}_{\frac{1}{5}, \frac{0}{5}}(5 \tau) \, \mathfrak{k}^{}_{\frac{2}{5}, \frac{0}{5}}(5 \tau) \; , & ~\quad & \widehat{e}^{}_{6}(\tau) = \dfrac{\eta^{15}(5 \tau)}{\eta^{3}(\tau)} \, \mathfrak{k}^{5}_{\frac{1}{5}, \frac{0}{5}}(5 \tau) \; .
\end{array}
\label{eq:basvec}
\end{eqnarray}
Furthermore, making use of Eqs.~(\ref{eq:Dedekindeta}) and (\ref{eq:Kexpansion}), we can derive the $q$-expansions of the above six basis vectors
\begin{eqnarray}
\widehat{e}^{}_1 & = & 1 + 3q + 4q^2_{} + 2q^3_{} + q^4_{} + 3q^5_{} + 6q^6_{} + 4q^7_{} - q^9_{} + \cdots \; , \nonumber \\
\widehat{e}^{}_2 & = & q^{1/5}_{} \left( 1 + 2q + 2q^2_{} + q^3_{} + 2q^4_{} + 2q^5_{} + 2q^6_{} + q^7_{} + 2q^8_{} + 2q^9_{} + \cdots \right) \nonumber \; , \\
\widehat{e}^{}_3 & = & q^{2/5}_{} \left( 1 + q + q^2_{} + q^3_{} + 2q^4_{} + q^6_{} + q^7_{} + 2q^8_{} + q^9_{} + \cdots \right) \; ,\nonumber \\
\widehat{e}^{}_4 & = & q^{3/5}_{} \left( 1 + q^2_{} + q^3_{} + q^4_{} - q^5_{} + 2q^6_{} + 2q^8_{} + q^9_{} + \cdots \right) \; ,  \nonumber \\
\widehat{e}^{}_5 & = & q^{4/5}_{} \left( 1 - q + 2q^2_{} + 2q^6_{} - 2q^7_{} + 2q^8_{} + q^9_{} +\cdots \right) \; , \nonumber \\
\widehat{e}^{}_6 & = & q \left( 1 - 2q + 4q^2_{} - 3q^3_{} + q^4_{} + 2q^5_{} - 2q^6_{} + 3q^8_{} - 2q^9_{} + \cdots \right) \; .
\label{eq:basisq}
\end{eqnarray}
From Eq.~(\ref{eq:basvec}), one can immediately observe that $\widehat{e}^{}_i\widehat{e}^{}_{j} = \widehat{e}^{}_m\widehat{e}^{}_{n}$ exactly holds for $i+j = m+n$. These relations are very useful for the calculations of higher-weight modular forms, as we shall see in the next subsection. Under the transformation of $T$, we have
\begin{eqnarray}
\widehat{e}^{}_1 \rightarrow \widehat{e}^{}_1 \; , \quad  \widehat{e}^{}_2 \rightarrow \omega \, \widehat{e}^{}_2 \; , \quad   \widehat{e}^{}_3 \rightarrow \omega^2_{} \, \widehat{e}^{}_3 \; , \quad
\widehat{e}^{}_4 \rightarrow \omega^3_{} \, \widehat{e}^{}_4 \; , \quad  \widehat{e}^{}_5 \rightarrow \omega^4_{} \, \widehat{e}^{}_5 \; , \quad  \widehat{e}^{}_6 \rightarrow \widehat{e}^{}_6 \; ,
\label{eq:basisT}
\end{eqnarray}
with $\omega \equiv e^{2{\rm i}\pi/5}$, while under the transformation of $S$ we obtain
\begin{eqnarray}
\widehat{e}^{}_1 &\rightarrow& (-{\rm i}\tau) \dfrac{\sqrt{\phi}}{\sqrt[4]{5}} \left[ \dfrac{\phi^2_{}}{5} \widehat{e}^{}_1 + \phi \widehat{e}^{}_2 + 2\widehat{e}^{}_3 + 2\left(\phi-1\right) \widehat{e}^{}_4 + \left(\phi-1\right)^2_{} \widehat{e}^{}_5 + \dfrac{2\phi-3}{5} \widehat{e}^{}_6 \right] \; , \nonumber \\
\widehat{e}^{}_2 &\rightarrow& (-{\rm i}\tau) \dfrac{\sqrt{\phi-1}}{\sqrt[4]{5}} \left[ \dfrac{\phi^2_{}}{5} \widehat{e}^{}_1 + \dfrac{\sqrt{5}}{5} \widehat{e}^{}_2 - \dfrac{2\sqrt{5}}{5} \widehat{e}^{}_3 - \dfrac{2\sqrt{5}\phi}{5} \widehat{e}^{}_4 - \dfrac{\sqrt{5}\phi}{5} \widehat{e}^{}_5 - \dfrac{(\phi-1)}{5} \widehat{e}^{}_6 \right] \; ,\nonumber \\
\widehat{e}^{}_3 &\rightarrow & (-{\rm i}\tau) \dfrac{\sqrt{\phi}}{\sqrt[4]{5}} \left[ \dfrac{1}{5} \widehat{e}^{}_1 - \dfrac{\sqrt{5} \left(\phi-1\right)}{5} \widehat{e}^{}_2 - \dfrac{\sqrt{5}}{5} \widehat{e}^{}_3 + \dfrac{\sqrt{5} \left(\phi-1\right)}{5} \widehat{e}^{}_4 + \dfrac{\sqrt{5}}{5} \widehat{e}^{}_5 + \dfrac{(\phi-1)}{5} \widehat{e}^{}_6 \right] \; ,  \nonumber \\
\widehat{e}^{}_4 &\rightarrow& (-{\rm i}\tau) \dfrac{\sqrt{\phi-1}}{\sqrt[4]{5}} \left[ \dfrac{1}{5} \widehat{e}^{}_1 - \dfrac{\sqrt{5}\phi}{5} \widehat{e}^{}_2 + \dfrac{\sqrt{5}}{5} \widehat{e}^{}_3 + \dfrac{\sqrt{5}\phi}{5} \widehat{e}^{}_4 - \dfrac{\sqrt{5}}{5} \widehat{e}^{}_5 - \dfrac{\phi}{5} \widehat{e}^{}_6 \right] \; , \nonumber \\
\widehat{e}^{}_5 &\rightarrow& (-{\rm i}\tau) \dfrac{\sqrt{\phi}}{\sqrt[4]{5}} \left[ \dfrac{\left(\phi-1\right)^2_{}}{5} \widehat{e}^{}_1 - \dfrac{\sqrt{5}}{5} \widehat{e}^{}_2 + \dfrac{2\sqrt{5}}{5} \widehat{e}^{}_3 - \dfrac{2\sqrt{5}(\phi-1)}{5} \widehat{e}^{}_4 - \dfrac{\sqrt{5}(\phi-1)}{5} \widehat{e}^{}_5 + \dfrac{\phi}{5} \widehat{e}^{}_6 \right] \; , \nonumber \\
\widehat{e}^{}_6 &\rightarrow& (-{\rm i}\tau) \dfrac{\sqrt{\phi-1}}{\sqrt[4]{5}} \left[ \dfrac{(\phi-1)^2_{}}{5} \widehat{e}^{}_1-(\phi-1) \widehat{e}^{}_2 + 2\widehat{e}^{}_3 - 2\phi\widehat{e}^{}_4 + \phi^2_{} \widehat{e}^{}_5 - \dfrac{2\phi+1}{5} \widehat{e}^{}_6 \right] \; ,
\label{eq:basisS}
\end{eqnarray}
with $\phi \equiv (\sqrt{5} + 1)/2$. The derivation of Eq.~(\ref{eq:basisT}) is quite straightforward, since one can simply use the $T$ transformation properties of $\eta(\tau)$ and $\mathfrak{k}^{}_{r^{}_{1}, r^{}_{2}}(\tau)$ shown in Eq.~(\ref{eq:etaklein}). However, the derivation of Eq.~(\ref{eq:basisS}) would be very tedious if one strictly followed the $S$ transformation properties of $\eta(\tau)$ and $\mathfrak{k}^{}_{r^{}_{1}, r^{}_{2}}(\tau)$. Our strategy for such a derivation is as follows. First of all, we know that each function $\widehat{e}^{}_i$ will be transformed under $S$ into a linear combination of all six basis vectors with coefficients to be determined. In each linear combination, the $q$-expansions of $\widehat{e}^{}_i$ given in Eq.~(\ref{eq:basisq}) will be performed. On the other hand, given the transformation rules for $\eta(\tau)$ and $\mathfrak{k}^{}_{r^{}_{1}, r^{}_{2}}(\tau)$ under $S$ in Eq.~(\ref{eq:etaklein}), one can calculate $\widehat{e}^{}_i$ after the $S$ transformation by using Eq.~(\ref{eq:basvec}) and perform the $q$-expansions as well. Since the expressions derived in those two different ways should be equivalent to each other, we can extract the coefficients in the linear combinations by comparing the first few terms in the $q$-expansions. With those coefficients, we can obtain the final results in Eq.~(\ref{eq:basisS}).

Now that the transformations of $\widehat{e}^{}_i$ under $S$ and $T$ are known, we are ready to determine the transformation rule for the modular form $Y^{(1)}_{\widehat{\bf 6}}(\tau)$ of weight one in the irreducible representation $\widehat{\bf 6}$, which can be expressed in terms of the basis vectors of the modular space ${\cal M}^{}_1 \left[\Gamma(5)\right]$. More explicitly, the components of $Y^{(1)}_{\widehat{\bf 6}}(\tau)$ can be written as
\begin{eqnarray}
\left( Y^{(1)}_{\widehat{\bf 6}}\right)^{}_i = \sum^6_{j = 1} a^{}_{ij} \widehat{e}^{}_j \; ,
\end{eqnarray}
for $i = 1, 2, \cdots , 6$, where the coefficients $a^{}_{ij}$ (for $i,j=1,2,\cdots,6$) need to be calculated. Then, after applying Eq.~(\ref{eq:ST}) in the case of $k = 1$ and $\mathbf{r} = \widehat{\bf 6}$ and taking the transformation properties of $\widehat{e}^{}_i$ in Eqs.~(\ref{eq:basisT}) and (\ref{eq:basisS}) into consideration, we can find all the nonzero coefficients
\begin{eqnarray}
a^{}_{11}=1 \; , \quad a^{}_{16} = a^{}_{61}= -3 \; , \quad a^{}_{22} =a^{}_{55}=5\sqrt{2} \; , \quad  a^{}_{33}=a^{}_{44}=10 \; , \quad a^{}_{66}=-1 \; .
\end{eqnarray}
As a consequence, the explicit expressions of all the components of $Y^{(1)}_{\widehat{\bf 6}}$ can be obtained. For later convenience, we denote six components of $Y^{(1)}_{\widehat{\bf 6}}(\tau)$ as $Y^{}_i(\tau)$ (for $i = 1, 2, \cdots , 6$) and then have
\begin{eqnarray}
Y^{(1)}_{\widehat{\bf 6}} =
\left(\begin{matrix}
Y^{}_1 \\
Y^{}_2 \\
Y^{}_3 \\
Y^{}_4 \\
Y^{}_5 \\
Y^{}_6 \\
\end{matrix}\right)
=\left(\begin{matrix}
\widehat{e}^{}_{1}-3 \, \widehat{e}^{}_6 \\
5\sqrt{2} \, \widehat{e}^{}_2 \\
10 \, \widehat{e}^{}_3 \\
10 \, \widehat{e}^{}_4 \\
5\sqrt{2} \, \widehat{e}^{}_5 \\
-3 \, \widehat{e}^{}_1-\widehat{e}^{}_6 \\
\end{matrix}\right) \; ,
\label{eq:Y_16}
\end{eqnarray}
where the argument $\tau$ has been suppressed for all the relevant functions, and the exact formulas of $\widehat{e}^{}_i$ and their $q$-expansions are given in Eqs.~(\ref{eq:basvec}) and (\ref{eq:basisq}), respectively.

\subsection{Modular forms of higher weights}

As usual, the modular forms of higher weights can be constructed through the tensor products of those of lower weights. We start with the modular forms of weight two (i.e., $k = 2$), which can be generated by the tensor product of two modular forms of weight one, namely, $Y^{(1)}_{\widehat{\bf 6}} \otimes Y^{(1)}_{\widehat{\bf 6}}$. With the help of the decomposition rules of tensor products in Appendix~\ref{app:B}, we get all the nonzero modular forms with weight two as
\begin{eqnarray}
Y^{(2)}_{{\bf 3}, {\rm i}} &=& \left[Y^{(1)}_{\widehat{\bf 6}} \otimes Y^{(1)}_{\widehat{\bf 6}}\right]^{}_{{\bf 3}^{}_{{\rm s},1}} =
-3 \left(
\begin{array}{c}
\widehat{e}^{2}_1 - 36\, \widehat{e}_1 \widehat{e}^{}_6 - \widehat{e}_6^2 \\
5 \sqrt{2}\, \widehat{e}_2^{} (\widehat{e}^{}_1 - 3\, \widehat{e}_6^{}) \\
5 \sqrt{2}\, \widehat{e}_5^{} (3\, \widehat{e}^{}_1 + \widehat{e}_6^{}) \\
\end{array}
\right)
= -3 \left(
\begin{array}{c}
Y^2_1-3Y^{}_1Y^{}_6-Y^2_{6} \\
Y^{}_1 Y^{}_2 \\
-Y^{}_5 Y^{}_6 \\
\end{array}\right) \; , \nonumber \\
Y^{(2)}_{{\bf 3},{\rm ii}} &=& \left[Y^{(1)}_{\widehat{\bf 6}} \otimes Y^{(1)}_{\widehat{\bf 6}}\right]^{}_{{\bf 3}^{}_{{\rm s},2}} = -\dfrac{\sqrt{6}}{9}Y^{(2)}_{{\bf 3},{\rm i}} \; , \nonumber \\
Y^{(2)}_{{\bf 3}^{\prime}_{},{\rm i}} &=& \left[Y^{(1)}_{\widehat{\bf 6}} \otimes Y^{(1)}_{\widehat{\bf 6}}\right]^{}_{{\bf 3}^{\prime}_{{\rm s},1}} =
\sqrt{6} \left(
\begin{array}{c}
-\left( \widehat{e}_1^2 + 14\, \widehat{e}_1^{} \widehat{e}_6^{} - \widehat{e}_6^2 \right) \\
5\sqrt{2}\, \widehat{e}_3^{} (\widehat{e}_1^{} + 2\, \widehat{e}_6^{}) \\
5\sqrt{2}\, \widehat{e}_4^{} (2\, \widehat{e}_1^{} - \widehat{e}_6^{}) \\
\end{array}
\right)
= \dfrac{1}{2} \left(
\begin{array}{c}
\sqrt{6}\left(Y^2_1-2Y^{}_1Y^{}_6-Y^{2}_6\right) \\
-\sqrt{3} Y^{}_3 (Y^{}_1+Y^{}_6) \\
\sqrt{3} Y^{}_4 (Y^{}_1-Y^{}_6) \\
\end{array}\right) \; , \nonumber \\
Y^{(2)}_{{\bf 3}^{\prime}_{},{\rm ii}} &=& \left[Y^{(1)}_{\widehat{\bf 6}} \otimes Y^{(1)}_{\widehat{\bf 6}}\right]^{}_{{\bf 3}^{\prime}_{{\rm s},2}} = -\dfrac{4\sqrt{6}}{3}Y^{(2)}_{{\bf 3}^{\prime}_{},{\rm i}} \; , \nonumber \\
Y^{(2)}_{\bf 5} &=& \left[Y^{(1)}_{\widehat{\bf 6}} \otimes Y^{(1)}_{\widehat{\bf 6}}\right]^{}_{{\bf 5}^{}_{\rm s}}=
5\left(
\begin{array}{c}
 \sqrt{2} \left(\widehat{e}_1^2 + \widehat{e}_6^2\right) \\
-2 \sqrt{3}\, \widehat{e}_2^{} (\widehat{e}_1^{} + 7\, \widehat{e}_6^{}) \\
2 \sqrt{3}\, \widehat{e}_3^{} (4\, \widehat{e}_6^{} - 3\, \widehat{e}_1^{}) \\
-2 \sqrt{3}\, \widehat{e}_4^{} (4\, \widehat{e}_1^{} + 3\, \widehat{e}_6^{}) \\
2 \sqrt{3}\, \widehat{e}_5^{} (\widehat{e}_6^{} - 7\, \widehat{e}_1^{}) \\
\end{array}
\right)=\dfrac{1}{2}\left(
\begin{array}{c}
\sqrt{2}\left(Y^2_1+Y^2_6 \right) \\
2\sqrt{6}Y^{}_2 \left( 2Y^{}_1+Y^{}_6\right) \\
\sqrt{3}Y^{}_3\left( Y^{}_6-3Y^{}_1\right) \\
\sqrt{3}Y^{}_4 \left( Y^{}_1+3Y^{}_6\right) \\
2\sqrt{6}Y^{}_5 \left(2Y^{}_6-Y^{}_1\right) \\
\end{array}\right) \; .
\label{eq:Y2}
\end{eqnarray}
Some comments on the above modular forms are in order. First, the dimensionality of the modular space ${\cal M}^{}_k \left[\Gamma(5)\right]$ is $5k + 1$, implying eleven independent modular forms of $k = 2$, which we take as $Y^{(2)}_{\bf 3} \equiv Y^{(2)}_{{\bf 3}, {\rm i}}$, $Y^{(2)}_{{\bf 3}^{\prime}_{}} \equiv Y^{(2)}_{{\bf 3}^{\prime}_{}, {\rm i}}$ and $Y^{(2)}_{\bf 5}$. Second, substituting the $q$-expansions of $\widehat{e}^{}_i$ in Eq.~(\ref{eq:basisq}) into the expressions of those three modular forms, we find that they are consistent up to some overall factors with the results obtained in Ref.~\cite{Ding:2019xna} for the modular $A^{}_5$ group. This should be the case, as the modular forms of even weights for the modular $A^{}_5$ group coincide with those for the double covering group $A^\prime_5$.

Following a similar procedure, we can derive the modular forms of weights three, four, five and six. For weight three, there exist sixteen independent modular forms, transforming as the irreducible representations $\widehat{\bf 4}$, $\widehat{\bf 6}^{}_1$ and $\widehat{\bf 6}^{}_2$ of the modular $A^{\prime}_{5}$ group, which can be expressed as
\begin{eqnarray}
Y^{(3)}_{\widehat{\bf 4}} &=& \left[Y^{(1)}_{\widehat{\bf 6}} \otimes Y^{(2)}_{\bf 3}\right]^{}_{\widehat{\bf 4}} = -\dfrac{3\sqrt{30}}{10}\left(Y^{2}_{1}-4Y^{}_1 Y^{}_6 -Y^{2}_{6}\right) \left(
\begin{matrix}
-\sqrt{2}Y^{}_2 \\
\sqrt{3} Y^{}_3 \\
\sqrt{3} Y^{}_{4} \\
\sqrt{2} Y^{}_5 \\
\end{matrix}\right) \; , \nonumber \\
Y^{(3)}_{\widehat{\bf 6},1} &=& \left[Y^{(1)}_{\widehat{\bf 6}} \otimes Y^{(2)}_{\bf 3}\right]^{}_{\widehat{\bf 6},2} = -\dfrac{\sqrt{3}}{2}
\left(
\begin{array}{c}
5 Y_1^3-12 Y_1^{2}Y_6^{}-11 Y_1^{}Y_6^{2}-2 Y_6^3 \\
-2 Y_2^{} \left(Y_1^2-5 Y_1^{}Y_6^{}-2 Y_6^2\right) \\
Y_3^{} \left(Y_1^{}+Y^{}_6\right)\left(Y^{}_1+2Y^{}_6\right) \\
Y_4^{} \left(Y_1^{}-Y^{}_6\right)\left(2Y^{}_1-Y^{}_6\right) \\
2 Y_5^{} \left(2 Y_1^2-5 Y_1^{}Y_6^{}-Y_6^2\right) \\
2 Y_1^3 - 11 Y_1^2 Y_6^{} + 12 Y_1^{} Y_6^2 + 5 Y_6^3 \\
\end{array}
\right) \; , \nonumber \\
Y^{(3)}_{\widehat{\bf 6},2} &=& \left[Y^{(1)}_{\widehat{\bf 6}} \otimes Y^{(2)}_{\bf 3^{\prime}_{}}\right]^{}_{\widehat{\bf 6},2} = -\dfrac{\sqrt{3}}{2}
\left(
\begin{array}{c}
3 Y_1^3-9 Y_1^{2}Y_6^{}-Y_1^{}Y_6^{2}+Y_6^3 \\
-2 Y_2^{} \left(2 Y_1^2-2 Y_1^{}Y_6^{}-Y_6^2\right) \\
2 Y_1^2 Y_3^{} \\
2 Y_4^{} Y_6^2 \\
2 Y_5^{} \left(Y_1^2-2 Y_1^{}Y_6^{}-2 Y_6^2\right) \\
-Y_1^3-Y_1^{2}Y_6^{}+9 Y_1^{}Y_6^{2}+3 Y_6^3 \\
\end{array}
\right) \; .
\end{eqnarray}
For weight four, we have
\begin{eqnarray}
Y^{(4)}_{\bf 1} &=& \left[Y^{(1)}_{\widehat{\bf 6}} \otimes Y^{(3)}_{\widehat{\bf 6},2}\right]^{}_{\bf 1} = -\sqrt{2}\left(Y^4_1-3Y^{3}_{1}Y^{}_{6}-Y^2_{1}Y^2_{6}+3Y^{}_1Y^{3}_{6}+Y^{4}_{6}\right) \; , \nonumber \\
Y^{(4)}_{\bf 3} &=&  \left[Y^{(1)}_{\widehat{\bf 6}} \otimes Y^{(3)}_{\widehat{\bf 6},2}\right]^{}_{{\bf 3}^{}_{{\rm s},1}} = \frac{\sqrt{3}}{4}
\left(
\begin{array}{c}
\left(Y_1^2+Y_6^2\right)\left(7 Y_1^2-18  Y_1^{} Y_6^{}-7 Y_6^2\right) \\
Y_2^{} \left(13 Y_1^3-3 Y_1^{2} Y_6-29 Y_1^{} Y_6^2 -9
Y_6^3\right) \\
-Y_5^{} \left(9 Y_1^3-29 Y_1^2 Y_6^{} +3 Y_1^{} Y_6^2 +13
Y_6^3\right) \\
\end{array}
\right) \; , \nonumber \\
Y^{(4)}_{\bf 3^{\prime}_{}} &=&  \left[Y^{(1)}_{\widehat{\bf 6}} \otimes Y^{(3)}_{\widehat{\bf 6},2}\right]^{}_{{\bf 3}^{\prime}_{{\rm s},1}} = -\frac{1}{2}
\left(
\begin{array}{c}
\sqrt{2}\left(Y^2_1+Y^2_6\right)\left(4Y^2_1-11Y^{}_1Y^{}_6-4Y^2_6\right) \\
-Y_3^{} \left(Y^{}_1-2Y^{}_6\right)\left(7Y^2_1-3Y^{}_1 Y^{}_6 -2Y^2_6\right) \\
Y_4^{} \left(2Y^{}_1+Y^{}_6\right)\left(2Y^{2}_1-3Y^{}_1 Y^{}_6-7Y^2_6\right) \\
\end{array}
\right) \; , \nonumber \\
Y^{(4)}_{\bf 4} &=&  \left[Y^{(1)}_{\widehat{\bf 6}} \otimes Y^{(3)}_{\widehat{\bf 6},2}\right]^{}_{{\bf 4}^{}_{\rm a}} = \left(Y^2_1 - 4 Y^{}_1 Y^{}_6 - Y^2_6 \right) \left(
\begin{array}{c}
\sqrt{2} Y_2^{} (2 Y_1^{}+Y_6^{}) \\
Y_3^{} (2 Y_1^{}+Y_6^{}) \\
-Y_4^{} (Y_1^{}-2 Y_6^{}) \\
\sqrt{2} Y_5^{} (Y_1^{}-2 Y_6^{}) \\
\end{array}
\right) \; , \nonumber \\
Y^{(4)}_{{\bf 5},1} &=&  \left[Y^{(1)}_{\widehat{\bf 6}} \otimes Y^{(3)}_{\widehat{\bf 4}}\right]^{}_{{\bf 5},2} =-\dfrac{3\sqrt{30}}{10} \left(Y^2_1 - 4 Y^{}_1 Y^{}_6 - Y^2_6 \right)
\left(
\begin{array}{c}
\sqrt{3} (Y_1^{}-3 Y_6^{}) (3 Y_1^{}+Y_6^{}) \\
-Y_2^{} (5 Y_1^{}+Y_6^{}) \\
\sqrt{2} Y_3^{} (Y_1^{}-Y_6^{}) \\
\sqrt{2} Y_4^{} (Y_1^{}+Y_6^{}) \\
-Y_5^{} (Y_1^{}-5 Y_6^{}) \\
\end{array}
\right) \; , \nonumber
\label{eq:Y4_1}
\end{eqnarray}
\begin{eqnarray}
Y^{(4)}_{{\bf 5},2} &=&  \left[Y^{(1)}_{\widehat{\bf 6}} \otimes Y^{(3)}_{\widehat{\bf 6},2}\right]^{}_{{\bf 5}^{}_{{\rm a},1}} =  \dfrac{1}{4}
\left(
\begin{array}{c}
11 Y_1^4-60 Y_1^3 Y_6^{}+58 Y_1^2 Y_6^2+60 Y_1^{}
Y_6^3+11 Y_6^4 \\
\sqrt{3} Y_2^{} (Y_1^{}+Y_6^{}) \left(Y_1^2+8 Y_1^{} Y_6^{}+3
Y_6^2\right) \\
-\sqrt{6} Y^{}_3 \left(Y_1^3+3 Y_1^2 Y_6^{}-9 Y_1^{}
Y_6^2-3 Y_6^3\right) \\
\sqrt{6} Y_4^{} \left(3 Y_1^3-9 Y_1^2 Y_6^{}-3 Y_1^{}
Y_6^2+Y_6^3\right) \\
-\sqrt{3} Y_5^{} (Y_1^{}-Y_6^{}) \left(3 Y_1^2-8 Y_1^{}
Y_6^{}+Y_6^2\right) \\
\end{array}
\right) \; .
\label{eq:Y4_2}
\end{eqnarray}
For weight five, we have
\begin{eqnarray}
Y^{(5)}_{\widehat{\bf 2}} &=& \left[Y^{(1)}_{\widehat{\bf 6}} \otimes Y^{(4)}_{{\bf 3}^{\prime}_{}}\right]^{}_{\widehat{\bf 2}} = \dfrac{\sqrt{6}}{4}\left(Y^2_1- 4 Y^{}_1 Y^{}_6 -Y^2_6\right)^2_{} \left(
\begin{matrix}
Y^{}_3 \\
Y^{}_4
\end{matrix}\right) \; , \nonumber \\
Y^{(5)}_{\widehat{\bf 2}^{\prime}_{}} &=& \left[Y^{(1)}_{\widehat{\bf 6}} \otimes Y^{(4)}_{\bf 3}\right]^{}_{\widehat{\bf 2}^{\prime}_{}} = \dfrac{3}{4}\left(Y^2_1- 4 Y^{}_1 Y^{}_6 -Y^2_6\right) \left(
\begin{array}{c}
Y_2^{} \left(7 Y_1^2-3 Y_1^{} Y_6^{} -2 Y_6^2\right) \\
Y_5^{} \left(2 Y_1^2-3 Y_1^{} Y_6^{}-7 Y_6^2\right) \\
\end{array}
\right) \; , \nonumber \\
Y^{(5)}_{\widehat{\bf 4}} &=& \left[ Y^{(1)}_{\widehat{\bf 6}} \otimes Y^{(4)}_{{\bf 3}^{\prime}_{}}\right]^{}_{\widehat{\bf 4}} = \dfrac{\sqrt{10}}{4}\left(Y^2_1- 4 Y^{}_1 Y^{}_6 -Y^2_6\right)  \left(
\begin{array}{c}
\sqrt{2} Y_2^{} \left(Y_1^2-6 Y_1^{} Y_6^{}-2 Y_6^2\right) \\
\sqrt{3} Y_3^{} Y_6^{} (2 Y_1^{}+Y_6^{}) \\
\sqrt{3} Y_1^{} Y_4^{} (Y_1^{}-2 Y_6^{}) \\
\sqrt{2} Y_5^{} \left(2 Y_1^2-6 Y_1^{} Y_6^{}-Y_6^2\right) \\
\end{array}
\right) \; , \nonumber \\
Y^{(5)}_{\widehat{\bf 6},1} &=& \left[ Y^{(1)}_{\widehat{\bf 6}} \otimes Y^{(4)}_{\bf 1}\right]^{}_{\widehat{\bf 6}} = -\sqrt{2}\left( Y_1^4 - 3  Y_1^3 Y_6^{} -  Y_1^2 Y_6^2 + 3  Y_1^{} Y_6^3 + Y_6^4 \right)  \left(
\begin{array}{c}
Y^{}_1 \\
Y^{}_2 \\
Y^{}_3 \\
Y^{}_4 \\
Y^{}_5 \\
Y^{}_6
\end{array}
\right) \; , \nonumber \\
Y^{(5)}_{\widehat{\bf 6},2} &=& \left[ Y^{(1)}_{\widehat{\bf 6}} \otimes Y^{(4)}_{\bf 3}\right]^{}_{\widehat{\bf 6},1} = \dfrac{\sqrt{6}}{16}  \left(
\begin{array}{c}
41 Y_1^5-195 Y_1^4 Y_6^{}+214 Y_1^3 Y_6^2+66 Y_1^2
Y_6^3-127 Y_1 Y_6^4-39 Y_6^5 \\
2 Y_2^{} \left(40 Y_1^4-81 Y_1^3 Y_6^{}-49 Y_1^2
Y_6^2+33 Y_1^{} Y_6^3+13 Y_6^4\right) \\
-Y_3^{} \left(Y_1^4+6 Y_1^3 Y_6^{}+20 Y_1^2 Y_6^2-114
Y_1^{} Y_6^3-41 Y_6^4\right) \\
Y_4^{} \left(41 Y_1^4-114 Y_1^3 Y_6^{}-20 Y_1^2 Y_6^2+6
Y_1^{} Y_6^3-Y_6^4\right) \\
2 Y_5^{} \left(13 Y_1^4-33 Y_1^3 Y_6^{}-49 Y_1^2
Y_6^2+81 Y_1^{} Y_6^3+40 Y_6^4\right) \\
39 Y_1^5-127 Y_1^4 Y_6^{}-66 Y_1^3 Y_6^2+214 Y_1^2
Y_6^3+195 Y_1^{} Y_6^4+41 Y_6^5 \\
\end{array}
\right) \; , \nonumber \\
Y^{(5)}_{\widehat{\bf 6},3} &=& \left[ Y^{(1)}_{\widehat{\bf 6}} \otimes Y^{(4)}_{\bf 4}\right]^{}_{\widehat{\bf 6},2} = \dfrac{21\sqrt{5}}{40} \left(Y^2_1- 4 Y^{}_1 Y^{}_6 -Y^2_6\right) \left(
\begin{array}{c}
(3 Y_1-Y_6) (3 Y_1+Y_6) (3 Y_6-Y_1) \\
2 Y_2^{} (Y_1+Y_6) (2 Y_1+Y_6) \\
Y_3^{} \left(5 Y_1^2-6 Y_1 Y_6-3 Y_6^2\right) \\
Y_4^{} \left(3 Y_1^2-6 Y_1 Y_6-5 Y_6^2\right) \\
-2 Y_5^{} (Y_1-2 Y_6) (Y_1-Y_6) \\
(3 Y_1+Y_6)  (Y_1+3 Y_6) (3 Y_6-Y_1) \\
\end{array}
\right) \nonumber \; .
\label{eq:Y5}
\end{eqnarray}
For weight six, we have
\begin{eqnarray}
Y^{(6)}_{\bf 1} &=& \left[ Y^{(1)}_{\widehat{\bf 6}} \otimes Y^{(5)}_{\widehat{\bf 6},2}\right]^{}_{\bf 1} = -\dfrac{3}{8}\left(Y_1^2 + Y_6^2\right) \left(41 Y_1^4 - 198 Y_1^3 Y_6^{} + 154 Y_1^2 Y_6^2 +  198 Y_1^{} Y_6^3 + 41 Y_6^4  \right)\; , \nonumber \\
Y^{(6)}_{{\bf 3},1} &=& \left[ Y^{(1)}_{\widehat{\bf 6}} \otimes Y^{(5)}_{\widehat{\bf 2}^{\prime}_{}}\right]^{}_{\bf 3} =\dfrac{9\sqrt{2}}{16} \left(Y^2_1- 4 Y^{}_1 Y^{}_6 -Y^2_6\right) \left(
\begin{array}{c}
(Y_1-3 Y_6) (3 Y_1+Y_6) \left(3 Y_1^2-2 Y_1
Y_6-3 Y_6^2\right) \\
2 Y_2^{} \left(2 Y_1^3-9 Y_1 Y_6^2-3 Y_6^3\right) \\
2 Y_5^{} \left(3 Y_1^3-9 Y_1^2 Y_6+2 Y_6^3\right) \\
\end{array}
\right) \; , \nonumber
\label{eq:Y6_1}
\end{eqnarray}
\begin{eqnarray}
Y^{(6)}_{{\bf 3},2} &=& \left[ Y^{(1)}_{\widehat{\bf 6}} \otimes Y^{(5)}_{\widehat{\bf 6},1}\right]^{}_{{\bf 3}^{}_{{\rm s},1}} =3\sqrt{2} \left(Y_1^4 - 3 Y_1^3 Y_6^{} - Y_1^2 Y_6^2 + 3 Y_1^{} Y_6^3 + Y_6^4\right) \left(
\begin{array}{c}
Y_1^2-3 Y_1^{} Y_6^{}-Y_6^2 \\
Y_1^{} Y_2^{} \\
-Y_5^{} Y_6^{} \\
\end{array}
\right) \; , \nonumber \\
Y^{(6)}_{{\bf 3}^{\prime}_{},1} &=& \left[ Y^{(1)}_{\widehat{\bf 6}} \otimes Y^{(5)}_{\widehat{\bf 2}}\right]^{}_{{\bf 3}^{\prime}_{}} =-\dfrac{\sqrt{3}}{2} \left(Y^2_1- 4 Y^{}_1 Y^{}_6 -Y^2_6\right)^2_{} \left(
\begin{array}{c}
(3 Y_1+Y_6) (Y_1^{}-3 Y_6^{}) \\
\sqrt{2} Y_1^{} Y_3^{} \\
\sqrt{2} Y_4^{} Y^{}_6 \\
\end{array}
\right) \; , \nonumber \\
Y^{(6)}_{{\bf 3}^{\prime}_{},2} &=& \left[ Y^{(1)}_{\widehat{\bf 6}} \otimes Y^{(5)}_{\widehat{\bf 6},1}\right]^{}_{{\bf 3}^{\prime}_{{\rm s},1}} =-\dfrac{\sqrt{6}}{2}\left( Y_1^4 - 3  Y_1^3 Y_6^{} -  Y_1^2 Y_6^2 + 3  Y_1^{} Y_6^3 + Y_6^4 \right) \left(
\begin{array}{c}
\sqrt{2} \left(Y_1^2-2 Y_1^{} Y_6^{}-Y_6^2\right) \\
-Y_3^{} (Y_1^{}+Y_6^{}) \\
Y_4^{} (Y_1^{}-Y_6^{}) \\
\end{array}
\right) \; , \nonumber \\
Y^{(6)}_{{\bf 4},1} &=& \left[ Y^{(1)}_{\widehat{\bf 6}} \otimes Y^{(5)}_{\widehat{\bf 2}}\right]^{}_{\bf 4} =-\dfrac{3}{4} \left(Y^2_1- 4 Y^{}_1 Y^{}_6 -Y^2_6\right)^2_{} \left(
\begin{array}{c}
-\sqrt{2} Y_2^{} (3 Y_1+Y_6) \\
Y_3^{} (Y_1+Y_6) \\
Y_4^{} (Y_1-Y_6) \\
\sqrt{2} Y_5^{} (Y_1-3 Y_6) \\
\end{array}
\right) \; , \nonumber \\
Y^{(6)}_{{\bf 4},2} &=& \left[ Y^{(1)}_{\widehat{\bf 6}} \otimes Y^{(5)}_{\widehat{\bf 2}^{\prime}_{}}\right]^{}_{\bf 4} =-\dfrac{\sqrt{6}}{8} \left(Y^2_1- 4 Y^{}_1 Y^{}_6 -Y^2_6\right) \left(
\begin{array}{c}
\sqrt{2} Y_2^{} \left(Y_1^3+11 Y_1^2 Y_6^{}+19 Y_1^{}
Y_6^2+5 Y_6^3\right) \\
Y_3^{} \left(13 Y_1^3-31 Y_1^2 Y_6^{}-17 Y_1^{}
Y_6^2-Y_6^3\right) \\
Y_4^{} \left(Y_1^3-17 Y_1^2 Y_6^{}+31 Y_1^{} Y_6^2+13
Y_6^3\right) \\
\sqrt{2} Y_5^{} \left(5 Y_1^3-19 Y_1^2 Y_6^{} + 11 Y_1^{}
Y_6^2-Y_6^3\right) \\
\end{array}
\right) \; , \nonumber \\
Y^{(6)}_{{\bf 5},1} &=& \left[ Y^{(1)}_{\widehat{\bf 6}} \otimes Y^{(5)}_{\widehat{\bf 4}}\right]^{}_{{\bf 5},2} =\dfrac{\sqrt{10}}{8} \left(Y^2_1- 4 Y^{}_1 Y^{}_6 -Y^2_6\right) \left(
\begin{array}{c}
\sqrt{3} (Y_1^{}-3 Y_6^{}) (3 Y_1^{}+Y_6^{})
\left(Y_1^2+Y_6^2\right) \\
-2 Y_2^{} (2 Y_1^{}+Y_6^{}) \left(2 Y_1^2-3 Y_1^{}
Y_6^{}-Y_6^2\right) \\
\sqrt{2} Y_3^{} \left(Y_1^3+2 Y_1^2 Y_6^{}-11 Y_1^{}
Y_6^2-4 Y_6^3\right) \\
\sqrt{2} Y_4^{} \left(4 Y_1^3-11 Y_1^2 Y_6^{}-2 Y_1^{}
Y_6^2+Y_6^3\right) \\
2 Y_5^{} (Y_1^{}-2 Y_6^{}) \left(Y_1^2-3 Y_1^{} Y_6^{}-2
Y_6^2\right) \\
\end{array}
\right) \; , \nonumber \\
Y^{(6)}_{{\bf 5},2} &=& \left[ Y^{(1)}_{\widehat{\bf 6}} \otimes Y^{(5)}_{\widehat{\bf 6},1}\right]^{}_{{\bf 5}^{}_{\rm s}} =-\dfrac{\sqrt{2}}{2}\left(Y_1^4 - 3 Y_1^3 Y_6^{} - Y_1^2 Y_6^2 + 3 Y_1^{} Y_6^3 + Y_6^4\right) \left(
\begin{array}{c}
\sqrt{2} \left(Y_1^2+Y_6^2\right) \\
2 \sqrt{6} Y_2^{} (2 Y_1^{}+Y_6^{}) \\
-\sqrt{3} Y_3^{} (3 Y_1^{}-Y_6^{}) \\
\sqrt{3} Y_4^{} (Y_1^{}+3 Y_6^{}) \\
-2 \sqrt{6} Y_5^{} (Y_1^{}-2 Y_6^{}) \\
\end{array}
\right) \; . \nonumber
\label{eq:Y_6_2}
\end{eqnarray}
Although only part of the above modular forms will be implemented to build the concrete models of neutrino masses and flavor mixing in the present paper, a complete list of them up to the weight $k = 6$ should be useful for future works.

\section{Minimal Seesaw Model}\label{sec:msm}

\subsection{Simple viable scenarios}

As a practical application of the modular $A^\prime_5$ group explored in the previous section, we consider the MSM of neutrino masses and lepton flavor mixing, in which two right-handed neutrino singlets can be just assigned into the two-dimensional irreducible representation of $A^{\prime}_{5}$. As we have mentioned, two-dimensional representations do not exist for the original $A^{}_5$ group. Therefore, the MSM with two right-handed neutrinos is a well-motivated and economical scenario for the model building with the modular $A^\prime_5$ group.

\begin{table}[t!]
	\centering
	\caption{Summary of the charge assignments of the chiral superfields under the ${\rm SU(2)^{}_{\rm L}}$ gauge symmetry and the modular $A^{\prime}_{5}$ symmetry in our model, where the corresponding weights have been listed in the last row.}
	\vspace{0.5cm}
	\begin{tabular}{cccccccc}
		\toprule
		&$\widehat{L}$ & $\widehat{E}_1^{\rm C}$ & $\widehat{E}_2^{\rm C}$ & $\widehat{E}_3^{\rm C}$ & $\widehat{N}^{\rm C}_{}$ & $\widehat{H}^{}_{\rm u}$, $\widehat{H}^{}_{\rm d}$ &  \\
		\midrule
		$\rm{SU(2)^{}_L}$ & 2 & 1 & 1 & 1 & 1 & 2 &  \\
		
		$A_5^{\prime}$&$\bf{3}$ & $\bf{1}$ & $\bf{1}$ &$\bf{1}$ & $\bf{\widehat{2}^{\prime}}$ & $\bf{1}$ & \\
		
		$k_I$ &$-2$ & 4 &6 &8 & 3 & 0 & \\
		\bottomrule
	\end{tabular}
	\label{table:chargeassign}
\end{table}

Together with two right-handed neutrino singlets assigned into $\widehat{\bf 2}^{\prime}_{}$ of the modular $A^{\prime}_{5}$ group, the charge assignments of other chiral superfields under the ${\rm SU}(2)^{}_{\rm L}$ gauge symmetry and the flavor $A^\prime_5$ symmetry in our model have been summarized in Table~\ref{table:chargeassign}. After the weights and representations of all the superfields under $A^{\prime}_{5}$ are fixed, it is easy to determine the Yukawa couplings that have to be the modular forms of weights $k^{}_{Y} = k^{}_{I^{}_1} + k^{}_{I^{}_2} + \cdots + k^{}_{I^{}_n}$, where $k^{}_{I^{}_i}$ (for $i = 1, 2, \cdots, n$) are the weights of the superfields involved. As for lepton masses and flavor mixing, the gauge- and modular-invariant superpotentials read
\begin{eqnarray}
\mathcal{W}^{}_l&=& + {\gamma}^{}_1 \left[\left(\widehat{L} \widehat{E}_1^{\rm C} \right)^{}_{\bf 3}  Y^{\left(2\right)}_{\bf{3}} \right]^{}_{\bf 1}\widehat{H}_{\rm d}^{}+{\gamma}^{}_2 \left[\left(\widehat{L} \widehat{E}_2^{\rm C} \right)^{}_{\bf 3}  Y^{\left(4\right)}_{\bf{3}} \right]^{}_{\bf 1}\widehat{H}_{\rm d}^{} \nonumber\\
&& +{\gamma}^{}_3 \left[\left(\widehat{L} \widehat{E}_3^{\rm C} \right)^{}_{\bf 3}  Y^{\left(6\right)}_{{\bf 3},1} \right]^{}_{\bf 1}\widehat{H}_{\rm d}^{} + {\gamma}^{}_4 \left[\left(\widehat{L} \widehat{E}_3^{\rm C} \right)^{}_{\bf 3}  Y^{\left(6\right)}_{{\bf 3},2} \right]^{}_{\bf 1}\widehat{H}_{\rm d}^{} \; , \nonumber \\
\mathcal{W}^{}_{\rm D}&=&g\left[\left(\widehat{L} \widehat{N}^{\rm C}_{} \right)^{}_{\bf \widehat{6}} Y^{\left(1\right)}_{\bf \widehat{6}} \right]_{\bf 1} \widehat{H}_{\rm u}^{} \; , \nonumber \\
\mathcal{W}^{}_{\rm R}&=& \dfrac{1}{2}\Lambda^{}_1 \left[\left(\widehat{N}^{\rm C}_{} \widehat{N}^{\rm C}_{}\right)^{}_{\bf 3^{\prime}} Y^{(6)}_{{\bf 3^{\prime}_{}},1} \right]^{}_{\bf 1}+\dfrac{1}{2}\Lambda^{}_2 \left[\left(\widehat{N}^{\rm C}_{} \widehat{N}^{\rm C}_{}\right)^{}_{\bf 3^{\prime}_{}} Y^{(6)}_{{\bf 3^{\prime}_{}},2} \right]^{}_{\bf 1} \; .
\label{eq:superp}
\end{eqnarray}
Without loss of generality, it is always possible to render $\gamma^{}_i$ (for $i = 1, 2, 3$), $g$ and $\Lambda^{}_1$ to be real by redefining the unphysical phases of lepton fields. Hence we are left with two complex parameters $\gamma^{}_{4}/\gamma^{}_{3} \equiv \widetilde{\gamma} = \gamma e^{{\rm i}\varphi^{}_{\gamma}}$ and $\Lambda^{}_{2}/\Lambda^{}_{1} \equiv \widetilde{\Lambda} = \Lambda e^{{\rm i}\varphi^{}_{\Lambda}}$. After implementing the Kronecker product rules in Appendix~\ref{app:B}, we obtain the charged-lepton mass matrix $M^{}_l$, the Dirac neutrino mass matrix $M^{}_{\rm D}$ and Majorana neutrino mass matrix $M^{}_{\rm R}$, i.e.,
\begin{eqnarray}
M_{l}&=&
\dfrac{v^{}_{\rm d}}{\sqrt{2}}\left(
\begin{matrix}
\left(Y^{(2)}_{\bf 3}\right)^{}_1 && \left(Y^{(4)}_{\bf 3}\right)^{}_1 && \left(Y^{(6)}_{{\bf 3},1}\right)^{}_1 + \widetilde{\gamma} \left(Y^{(6)}_{{\bf 3},2}\right)^{}_1  \\
\left(Y^{(2)}_{\bf 3}\right)^{}_3 &&\left(Y^{(4)}_{\bf 3}\right)^{}_3 && \left(Y^{(6)}_{{\bf 3},1}\right)^{}_3+\widetilde{\gamma} \left(Y^{(6)}_{{\bf 3},2}\right)^{}_3  \\
\left(Y^{(2)}_{\bf 3}\right)^{}_2 &&\left(Y^{(4)}_{\bf 3}\right)^{}_2 && \left(Y^{(6)}_{{\bf 3},1}\right)^{}_2+\widetilde{\gamma} \left(Y^{(6)}_{{\bf 3},2}\right)^{}_2  \\
\end{matrix}
\right)_{}^{*} \cdot
\left(
\begin{matrix}
\gamma^{}_1 && 0 && 0  \\
0 && \gamma^{}_2 && 0  \\
0 && 0 && \gamma^{}_3  \\
\end{matrix}
\right), \nonumber \\
M_{\rm D} &=& \dfrac{g v_{\rm u}^{}}{\sqrt{2}}
\left(
\begin{array}{cc}
\left(Y^{(1)}_{\bf{\widehat{6}}}\right)^{}_5&\left(Y^{(1)}_{\bf{\widehat{6}}}\right)^{}_2\\
-\left(Y^{(1)}_{\bf{\widehat{6}}}\right)^{}_4&-\left(Y^{(1)}_{\bf{\widehat{6}}}\right)^{}_1\\
-\left(Y^{(1)}_{\bf{\widehat{6}}}\right)^{}_6&\left(Y^{(1)}_{\bf{\widehat{6}}}\right)^{}_3\\
\end{array}
\right)_{}^{*}, \nonumber \\
M_{\rm R}&=& \Lambda^{}_1
\left(
\begin{array}{cc}
-\sqrt{2}\left[\left(Y^{(6)}_{{\bf 3^{\prime}},1} \right)^{}_3 + \widetilde{\Lambda} \left(Y^{(6)}_{{\bf 3^{\prime}}_{},2} \right)^{}_3  \right] & \left(Y^{(6)}_{{\bf 3}^{\prime}_{},1} \right)^{}_1 + \widetilde{\Lambda}  \left(Y^{(6)}_{{\bf 3}^{\prime}_{},2} \right)^{}_1 \\
\left(Y^{(6)}_{{\bf 3^{\prime}_{}},1} \right)^{}_1 + \widetilde{\Lambda}  \left(Y^{(6)}_{{\bf 3^{\prime}_{}},2} \right)^{}_1&\sqrt{2}\left[\left(Y^{(6)}_{{\bf 3^{\prime}_{}},1} \right)^{}_2 + \widetilde{\Lambda} \left(Y^{(6)}_{{\bf 3^{\prime}_{}},2} \right)^{}_2  \right]
\end{array}
\right)_{}^{*}, \label{eq:matrices}
\end{eqnarray}
where the asterisk ``$*$'' indicates the complex conjugation, $(Y^{(k)}_{\bf r})^{}_{i}$ denotes the $i$-th component of $Y^{(k)}_{\bf r}$, $v^{}_{\rm u}$ and $v^{}_{\rm d}$ stand for the vev's of the up- and down-type Higgs fields in the minimal supersymmetric standard model (MSSM). Once $M^{}_{\rm D}$ and $M^{}_{\rm R}$ are known, we shall be able to get the neutrino mass matrix $M^{}_{\nu}$ via the seesaw formula $M^{}_{\nu} = -M^{}_{\rm D} M^{-1}_{\rm R} M^{\rm T}_{\rm D}$. With both $M^{}_l$ and $M^{}_\nu$, one can diagonalize them and get lepton masses and flavor mixing matrix.

First, we carry out a detailed numerical analysis of our model, for which a parameter counting should be helpful. In addition to the real and imaginary parts $\{{\rm Re}\,\tau, {\rm Im}\,\tau\}$ of the modulus $\tau$, there are five parameters in the charged-lepton sector (i.e., $\mu^{}_l \equiv v^{}_{\rm d}\gamma^{}_3/\sqrt{2} $, $b^{}_1 \equiv \gamma^{}_1/\gamma^{}_3$, $b^{}_2 \equiv \gamma^{}_2/\gamma^{}_3$ and $\widetilde{\gamma} \equiv \gamma e^{{\rm i}\varphi^{}_{\gamma}}$) and three parameters in the neutrino sector (i.e., $\mu^{}_{\nu} \equiv g^{2}_{}v^{2}_{\rm u}/(2\Lambda^{}_{1})$ and $\widetilde{\Lambda} \equiv \Lambda e^{{\rm i}\varphi^{}_{\Lambda}}$). Totally, we have ten real model parameters. The most general case, where all the ten parameters are taken into account, will be numerically studied. Then, we assume two of the free parameters to be zero and investigate two such special scenarios. The strategy for our numerical analysis is quite simple. Upon scanning over the parameter space of the model parameters, we can compute the low-energy observables, namely, two neutrino mass-squared differences $\{\Delta m^{2}_{21}, \Delta m^{2}_{31} ~{\rm or}~\Delta m^{2}_{32}\}$ and three flavor mixing angles $\{\theta^{}_{12}, \theta^{}_{13}, \theta^{}_{23}\}$, which are then confronted with their allowed ranges at the $1\sigma$ or $3\sigma$ level according to the global-fit results from NuFIT 5.0~\cite{Esteban:2020cvm,nufit5.0} without including the atmospheric neutrino data from Super-Kamiokande, as summarized in Table~\ref{table:gfit}. Then the CP-violating phases and effective neutrino masses for beta decays and neutrinoless double beta decays are calculated as theoretical predictions. More details of our numerical calculations and comments on the final results can be found below.
\begin{itemize}
\item The values of $\tau$ (i.e., both real and imaginary parts) are randomly generated from the region
	\begin{eqnarray}
	{\cal G}^{}_{\rm R} = \left\{ \tau \in \mathbb{C}: \quad {\rm Im}\,\tau > 0, \;  0 \leq {\rm Re}\,\tau \leq 0.5, \; |\tau| \geq 1 \right\} \; ,
	\label{eq:fundo}
	\end{eqnarray}
	which is the right-half part of the fundamental domain ${\cal G}$. As for the conjugate part with $-0.5 \leq {\rm Re}\,\tau \leq 0$, the corresponding results can be obtained by reversing the signs of all the phases. Moreover, $\gamma$ and $\varphi^{}_{\gamma}$ are chosen from the regions $\gamma \in [10^{-4}_{},10^{4}_{}]$ and $\varphi^{}_{\gamma} \in [0,2\pi)$, respectively. Then three real parameters $\mu^{}_l$, $b^{}_1$ and $b^{}_2$ in the charged-lepton sector can be numerically determined via the following identities
\begin{eqnarray}
	{\rm Tr} \left( M^{}_l M^{\dag}_l \right) &=& m^{2}_{e} + m^{2}_{\mu} + m^{2}_{\tau} \; ,  \label{eq:tr}\\
	{\rm Det}\left( M^{}_l M^{\dag}_l \right) &=& m^{2}_{e} m^{2}_{\mu} m^{2}_{\tau} \; , \label{eq:det}\\
	\dfrac{1}{2}\left[{\rm Tr} \left(M^{}_l M^{\dag}_l\right)\right]^2_{} - \dfrac{1}{2}{\rm Tr}\left[ (M^{}_l M^{\dag}_l)^2_{}\right] &= & m^{2}_{e}m^{2}_{\mu}+m^{2}_{\mu}m^{2}_{\tau}+m^{2}_{\tau}m^{2}_{e} \; , \label{eq:tr2}
	\end{eqnarray}
where the running charged-lepton masses $m^{}_e = 0.48307~{\rm MeV}$, $m^{}_{\mu} = 0.101766~{\rm GeV}$ and $m^{}_{\tau} = 1.72856~{\rm GeV}$ are evaluated at the electroweak scale characterized by $m^{}_{Z} = 91.2~{\rm GeV}$~\cite{Huang:2020hdv}.\footnote{Although the modular symmetry is usually supposed to work at some high-energy scale, the renormalization-group running effects can be safely neglected in our model for two reasons. First, neutrino masses cannot be nearly-degenerate in the MSM, where the lightest neutrino turns out to be massless. Second, a sufficiently small value of $\tan\beta = v^{}_{\rm u}/v^{}_{\rm d}$ is assumed such that the charged-lepton Yukawa couplings are highly suppressed.} After so doing, the charged-lepton mass matrix $M^{}_l$ is also fixed, and the unitary matrix $U^{}_{l}$ used to diagonalize it via $U^\dagger_l M^{}_l M^\dagger_l U^{}_l = {\rm Diag}\{m^2_e, m^2_\mu, m^2_\tau\}$ can be obtained.
	
\item In the neutrino sector, we randomly generate the values of $\Lambda$ and $\varphi^{}_{\Lambda}$ in the regions $\Lambda \in [10^{-4}_{},10^4]$ and $\varphi^{}_{\Lambda} \in [0,2\pi)$, respectively. Given the modulus parameter $\tau$ in the previous step, the neutrino mass matrix $M^{}_\nu$ can be determined up to the overall factor $\mu^{}_{\nu}$. Therefore we can calculate the ratio $r^{}_{} \equiv \sqrt{\Delta m^{2}_{21}/\Delta m^{2}_{31}} = m^{}_2/m^{}_3$ in the case of normal mass ordering (NO) with $m^{}_1 = 0 < m^{}_2 < m^{}_3$ or $r^{}_{} \equiv \sqrt{\Delta m^{2}_{21}/|\Delta m^{2}_{32}|} = \sqrt{1 - m^2_1/m^2_2}$ in the case of inverted mass ordering (IO) with $m^{}_3 = 0 < m^{}_1 < m^{}_2$. This ratio is independent of the overall neutrino mass scale $\mu^{}_{\nu}$, so is the unitary matrix $U^{}_{\nu}$ diagonalizing the neutrino mass matrix $M^{}_{\nu}$ via $U^{\dag}_{\nu} M^{}_{\nu} M^{\dag}_{\nu} U^{}_{\nu}= {\rm Diag}\left\{m^{2}_{1}, m^{2}_{2}, m^{2}_{3}\right\}$. The lepton flavor mixing matrix $U$, or the Pontecorvo-Maki-Nakagawa-Sakata (PMNS) matrix~\cite{Pontecorvo:1957cp,Maki:1962mu}, is thus given by $U = U^{\dag}_{l}U^{}_{\nu}$. In the standard parametrization of the PMNS matrix~\cite{PDG2020}, we have
	\begin{eqnarray}
	U = \left(
	\begin{matrix}
	c^{}_{12}c^{}_{13} && s^{}_{12}c^{}_{13} && s^{}_{13}e^{-{\rm i}\delta}_{} \\
	-s^{}_{12}c^{}_{23}-c^{}_{12}s^{}_{23}s^{}_{13}e^{{\rm i}\delta} && c^{}_{12}c^{}_{23}-s^{}_{12}s^{}_{23}s^{}_{13}e^{{\rm i}\delta}_{} && s^{}_{23}c^{}_{13} \\
	s^{}_{12}s^{}_{23}-c^{}_{12}c^{}_{23}s^{}_{13}e^{{\rm i}\delta} && -c^{}_{12}s^{}_{23}-s^{}_{12}c^{}_{23}s^{}_{13}e^{{\rm i}\delta} &&
	c^{}_{23}c^{}_{13}
	\end{matrix}\right) \cdot \left(
	\begin{matrix}
	1 && ~  && ~\\
	~ && e^{{\rm i}\sigma} && ~ \\
	~ && ~ && 1
	\end{matrix}\right) \; ,
	\label{eq:UPMNS}
	\end{eqnarray}
	where $c^{}_{ij} \equiv \cos \theta^{}_{ij}$ and $s^{}_{ij} \equiv \sin \theta^{}_{ij}$ (for $ij =12,13,23$) have been defined. Note that in the MSM, the lightest neutrino is massless and thus only one Majorana CP-violating phase $\sigma$ is physical. Comparing the obtained values of $r^{}_{}$ and $\{\sin^2\theta^{}_{12}, \sin^2\theta^{}_{13}, \sin^2\theta^{}_{23}\}$ with their allowed $3\sigma$ (or $1\sigma$) ranges from global-fit results, we find out the parameter space of our model that is compatible with experimental data at the $3\sigma$ (or $1\sigma$) level. The overall factor $\mu^{}_{\nu}$ can be pinned down by further reproducing the correct values of $\Delta m^2_{21}$ or $|\Delta m^2_{31}|$, but it is not relevant for our discussions about lepton flavor mixing and CP violation.

\item In order to determine the best-fit values of free parameters in our model, we construct the $\chi^2$-function as the sum of one-dimensional functions $\chi^2_{j}$, namely,
	\begin{eqnarray}
	\chi^2_{}(p^{}_{i}) = \sum^{}_{j}\chi^2_{j}(p^{}_i)\; ,
	\label{eq:chi}
	\end{eqnarray}
	where $p^{}_{i} \in \{{\rm Re}\,\tau, {\rm Im}\,\tau, \gamma, \varphi^{}_{\gamma}, \Lambda, \varphi^{}_{\Lambda}\}$ stand for the model parameters, and $j$ is summed over the observables $\{\sin^2\theta^{}_{12}, \sin^2\theta^{}_{13}, \sin^2\theta^{}_{23}, r\}$. For $\sin^2\theta^{}_{12}$, $\sin^2\theta^{}_{13}$ and $r$, we simply take the Gaussian approximations and assume
	\begin{eqnarray}
	\chi^2_{j}(p^{}_{i}) = \left(\frac{q^{}_{j}(p^{}_{i})-q^{\rm bf}_{j}}{\sigma^{}_{j}}\right)^2_{} \; ,
	\label{eq:ch2}
	\end{eqnarray}
	where $q^{}_j(p^{}_i)$ denote the model predictions for these observables and $q^{\rm bf}_j$ are their best-fit values from the global analysis in Ref.~\cite{Esteban:2020cvm}. The associated uncertainties $\sigma^{}_j$ are derived by symmetrizing $1\sigma$ uncertainties from the global-fit analysis, and have already been given in Table~\ref{table:gfit}. For $\sin^2_{}\theta^{}_{23}$, we use the one-dimensional projection of the $\chi^2$-function provided by Refs.~\cite{Esteban:2020cvm,nufit5.0}. By minimizing the overall $\chi^2_{}$-function in Eq.~(\ref{eq:chi}), we can determine the best-fit values of our model parameters.
\end{itemize}

\begin{figure}[t!]
	\centering		\includegraphics[width=0.56\textwidth]{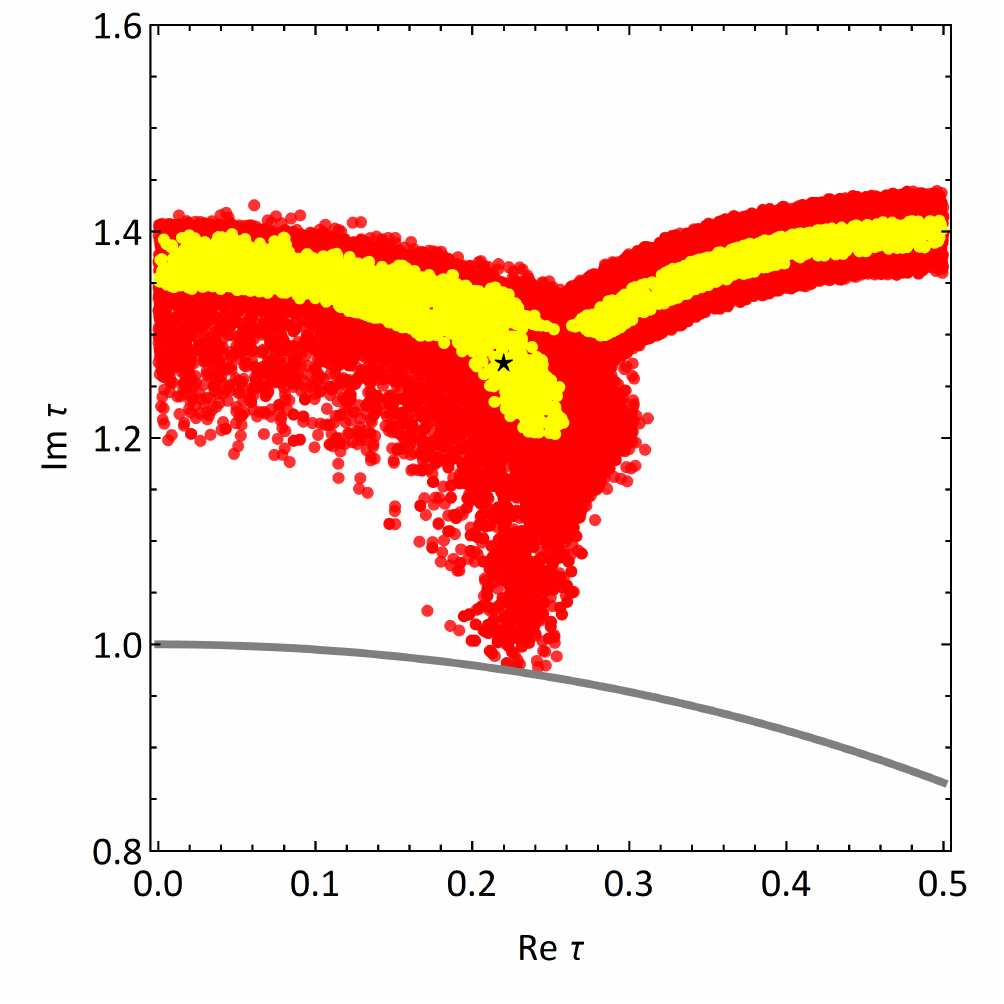}
	\vspace{-0cm}
	\caption{The allowed parameter space of the model parameters $\{{\rm Re}\,\tau, {\rm Im}\,\tau\}$ in the most general case with ten real parameters, where the $1\sigma$ (yellow dots) and $3\sigma$ (red dots) ranges of flavor mixing angles and neutrino mass-squared differences from the global-fit analysis of neutrino oscillation data have been input~\cite{Esteban:2020cvm}. The best-fit values of $\{{\rm Re}\,\tau, {\rm Im}\,\tau\}$ from the $\chi^2_{}$-fit analysis are denoted by the black star. In addition, the lower boundary of ${\cal G}^{}_{\rm R}$ is plotted as the gray curve.}
	\label{fig:MSM_All} 
\end{figure}

\begin{table}[t!]
	\begin{center}
		\vspace{0cm} \caption{Summary of the best-fit values with $\chi^2_{\rm min} = 0.0153$, together with the 1$\sigma$ and 3$\sigma$ ranges, of all the free model parameters in the NO case in our model with ten real parameters. The allowed ranges of the low-energy observables are also given.} \vspace{0.5cm}
		\begin{tabular}{c|c|c|c|c}
			\hline
			\hline
			\multicolumn{2}{c|}{} & Best fit & $1\sigma$ range & $3\sigma$ range \\
			\hline
			\multirow{10}*{\rotatebox[]{90}{Free model parameters}}
			& ${\rm Re}\,\tau$ & 0.2201 & 0 -- 0.5 & 0 -- 0.5\\
			~ & ${\rm Im}\,\tau$ & 1.272 & 1.204 -- 1.410 & 0.978 -- 1.439 \\
			~ & $\gamma$              & 13.51 & 0 -- 570 & 0 -- 1293 \\
			~ & $\varphi^{}_{\gamma}$ $/{}^{\circ}_{}$ & 346.8 & 0 -- 360 & 0 -- 360 \\
			~ & $\Lambda$ & 5.606 & 1.061 -- 17.17 & 0 -- 19.90  \\
			~ & $\varphi^{}_{\Lambda}/ {}^{\circ}_{}$ & 209.3 & 82.47 -- 317.0 & 0 -- 360 \\
			~ & $\mu^{}_l/{\rm MeV}$ &  0.1369 & $0.1158$ -- $150.5$ & $5.163 \times 10^{-2}_{}$ -- $1.789 \times 10^{2}_{}$ \\
			~ & $b^{}_1$ & 2413 & $9.243 \times 10^{-3}_{}$ -- $2.771 \times 10^3_{}$ & $2.174 \times 10^{-3}_{}$  -- $5.301 \times 10^3_{}$ \\
			~ & $b^{}_2$ & 85.60 & $1.235 \times 10^{-3}$ -- $ 1.361 \times 10^2_{}$ & $5.210 \times 10^{-4}$ -- $ 5.713 \times 10^2_{}$ \\
			~ & $\mu^{}_{\nu}/{\rm meV}$ & 26.60 & 24.11 -- 48.93 & 23.68 -- 87.16\\
			\hline
			\multirow{9}*{\rotatebox[]{90}{Observables}}
			~ & $m^{}_2/{\rm meV}$& 8.614 & 8.497 -- 8.735 & 8.258 -- 8.967\\
			~ & $m^{}_3/{\rm meV}$& 50.19 & 49.87 -- 50.42 & 49.30 -- 50.98 \\
			~ & $\theta^{}_{12}/{}^{\circ}_{}$ & 33.51 & 32.70 -- 34.21 & 31.27 -- 35.86\\
			~ & $\theta^{}_{13}/{}^{\circ}_{}$ & 8.582 & 8.45 -- 8.70 & 8.20 -- 8.97 \\
			~ & $\theta^{}_{23}/{}^{\circ}_{}$& 49.02 & 47.6 -- 50.1 & 39.6 -- 51.8\\
			~ & $\delta/{}^{\circ}_{}$& 359.6  & 0 -- 360 & 0 -- 360\\
			~ & $\sigma/{}^{\circ}_{}$& 128.2 & 0 -- 180 & 0 -- 180\\
			~ & $m^{}_{\beta}/{\rm meV}$& 8.816 & 8.711 -- 8.943 & 8.259 -- 9.445 \\
			~ & $m^{}_{\beta\beta}/{\rm meV}$ & 2.404 & 1.428 -- 3.676 & 1.043 -- 4.167\\
			\hline
			& $\chi^2_{\rm min}$ & 0.0153 &  -- & -- \\
			\hline
			\hline
		\end{tabular}
		\label{table:value}
	\end{center}
\vspace{-0.25cm}
\end{table}

In the most general case, where all the ten real parameters are included, our model is compatible with the global-fit results of oscillation parameters at the $1\sigma$ level in the NO case, while the IO case is excluded at the $3\sigma$ level. The allowed parameter space of $\{{\rm Re}\,\tau, {\rm Im}\,\tau\}$ in the NO case is shown in Fig.~\ref{fig:MSM_All}, where one can observe that the whole range $[0,0.5]$ of ${\rm Re}\,\tau$ is allowed, while ${\rm Im}\,\tau$ can vary from 0.98 to 1.44 at the $3\sigma$ level. In Table~\ref{table:value}, we summarize the best-fit values, as well as the $1\sigma$ and $3\sigma$ ranges, of the free parameters, and the allowed ranges of low-energy observables. Notice that the parameter $\gamma$ can vary in a broad range from 0 to 1293. It is interesting to see that $\gamma^{}_3$ could be much smaller than $\gamma^{}_1$ and $\gamma^{}_2$, so the contribution to the flavor mixing from the third column of $M^{}_l$ can be negligible when compared to those from the first two columns. On the other hand, Table~\ref{table:value} reveals that $\widetilde{\gamma}$ or $\widetilde{\Lambda}$ could even be zero, providing the possibility to reduce the number of free model parameters from ten to eight.

Next, motivated by the above observations, we now consider two special cases, namely, \emph{Case~A} with $\widetilde{\Lambda}=0$ and \emph{Case~B} with $\widetilde{\gamma}=0$.
\begin{figure}[t!]
	\centering		\includegraphics[width=0.98\textwidth]{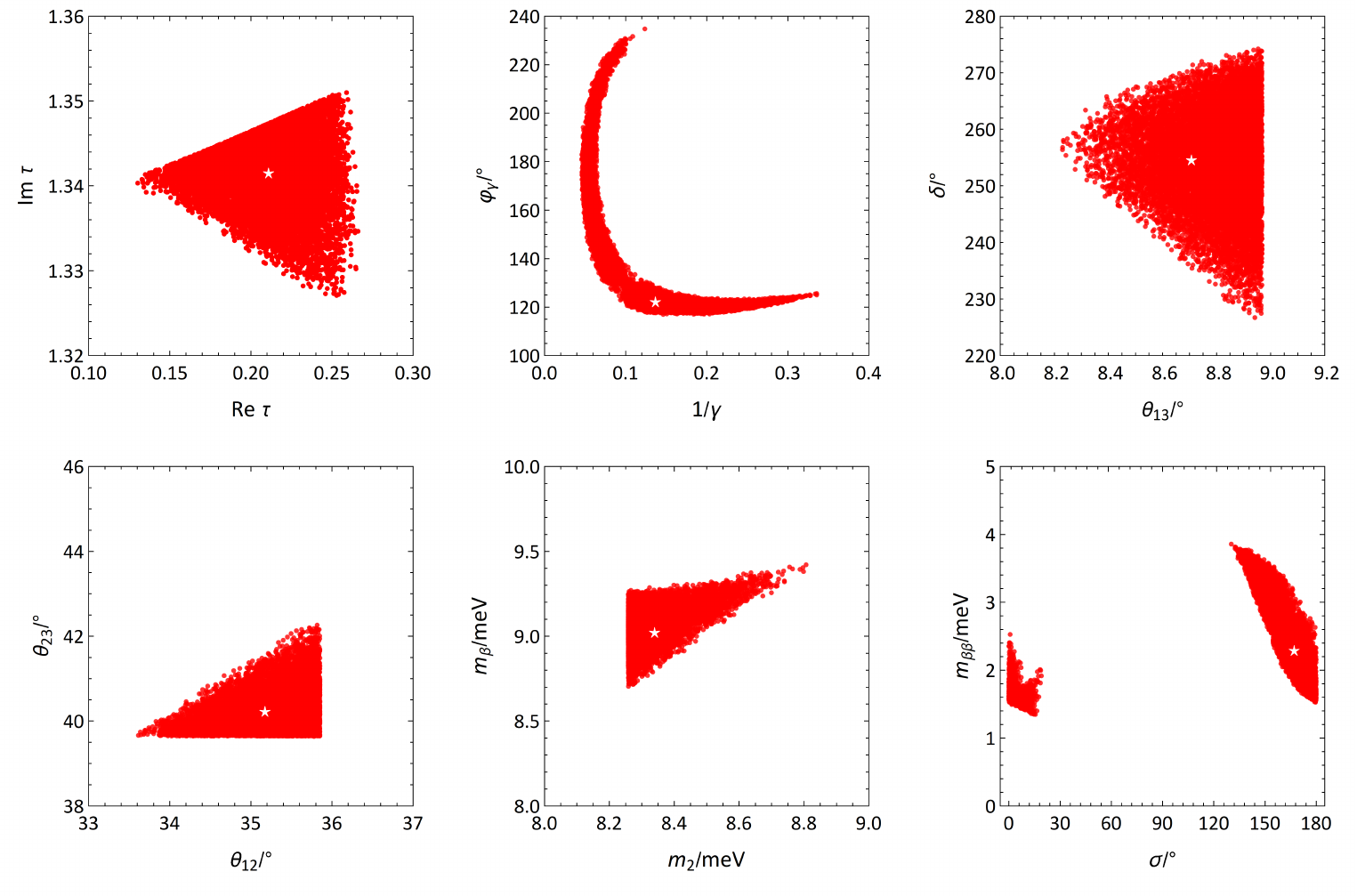}
	\vspace{-0.3cm}
	\caption{The allowed parameter space of model parameters, as well as the constraints on the low-energy observables in \emph{Case~A} with $\widetilde{\Lambda}=0$, where the $3\sigma$ ranges of flavor mixing angles and neutrino mass-squared differences from the global-fit analysis of neutrino oscillation data have been input~\cite{Esteban:2020cvm}. The best-fit values from the $\chi^2_{}$-fit analysis are denoted by the white stars.}
	\label{fig:MSM_R1} 
\end{figure}
\begin{itemize}
\item In \emph{Case~A} with $\widetilde{\Lambda}=0$, only $Y^{(6)}_{{\bf 3^{\prime}_{}},1}$ is retained in the right-handed neutrino mass matrix $M^{}_{\rm R}$. We find that this scenario is consistent with the global-fit analysis of neutrino oscillation data at the $3\sigma$ level in the NO case. The allowed parameter space and the constraint on low-energy observables are shown in Fig.~\ref{fig:MSM_R1}. From Fig.~\ref{fig:MSM_R1}, we observe that the allowed parameter space of $\tau$ is restricted to the region of $0.12 \lesssim {\rm Re}\,\tau \lesssim 0.27$ and $1.326 \lesssim {\rm Im}\,\tau \lesssim 1.352$. In addition, $1/\gamma$ varies from 0.045 to 0.33, or equivalently $ 3 \lesssim \gamma \lesssim 22$, and $115^{\circ}_{} \lesssim \phi^{}_{\gamma} \lesssim 235^{\circ}_{}$ is obtained. There exist strong correlations among the low-energy observables. As can be seen from the top-right panel, the allowed range of $\delta$ decreases as the value of $\theta^{}_{13}$ becomes smaller. In particular, when $\theta^{}_{13}$ is as small as $8.2^{\circ}_{}$, the value of $\delta$ is tightly restricted to be around $257^{\circ}_{}$. From the bottom-left panel, we see the correlation between $\theta^{}_{12}$ and $\theta^{}_{23}$, indicating that relatively large values of $\theta^{}_{12}$ and small values of $\theta^{}_{23}$ are predicted in \emph{Case~A}. Next-generation neutrino oscillation experiments, e.g., JUNO~\cite{An:2015jdp}, Hyper-Kamiokande~\cite{Abe:2018uyc} and DUNE~\cite{Abi:2020evt}, will unambiguously pin down the neutrino mass ordering and determine the octant of $\theta^{}_{23}$, so \emph{Case A} will be hopefully confirmed or ruled out in the near future. Notice that $m^{}_1 = 0$ in the NO case, so the absolute neutrino masses can be immediately determined from neutrino mass-squared differences, namely, $m^{}_2 = \sqrt{\Delta m^2_{21}}$ and $m^{}_3 = \sqrt{\Delta m^2_{31}}$. The value of $m^{}_2$ cannot reach its upper bound of the $3\sigma$ allowed range from neutrino oscillation data. For the maximum $m^{}_2 = 8.81~{\rm meV}$, the effective mass for beta decays $m^{}_{\beta}\equiv \sqrt{m^2_1 |U^{}_{e1}|^2 + m^2_2 |U^{}_{e2}|^2 + m^2_3 |U^{}_{e3}|^2}$ is constrained to be $9.45~{\rm meV}$. Meanwhile, the effective mass for neutrinoless double-beta decays $m^{}_{\beta\beta}\equiv |m^{}_1 U^2_{e1} + m^{}_2 U^2_{e2} + m^{}_3 U^2_{e3}|$ is lying in the range $(1.35 \cdots 3.86)~{\rm meV}$, which will be a great challenge for future neutrinoless double-beta decay experiments~\cite{Cao:2019hli, Dolinski:2019nrj}.

   Based on the $\chi^2$-fit analysis, we find that the minimum $\chi^{2}_{\rm min} = 15.99$ is obtained in the NO case with the following best-fit values of the model parameters
  \begin{eqnarray}
  {\rm Re}\,\tau = 0.2110\;, \quad {\rm Im}\,\tau = 1.341 \;, \quad
  \gamma \ = 7.294 \;, \quad \varphi^{}_{\gamma} = 121.9^{\circ}_{} \;,
  \label{eq:bfcaseA}
  \end{eqnarray}
  which together with the charged-lepton masses lead to $\mu^{}_l = 0.03639~{\rm GeV}$, $b^{}_1 = 0.04809$ and $b^{}_2 = 0.3528$. The overall factor $\mu^{}_{\nu}$ in the neutrino sector  turns out to be $27.11~{\rm meV}$. Given the best-fit values of model parameters, we get the neutrino mass spectrum $m^{}_{1} = 0$, $m^{}_{2}=8.340~{\rm meV}$ and $m^{}_{3}=50.65~{\rm meV}$, three mixing angles $\theta^{}_{12}=35.17^{\circ}_{}$, $\theta^{}_{13}=8.707^{\circ}_{}$, $\theta^{}_{23} = 40.21^{\circ}_{}$ and two CP-violating phases $\delta=254.5^{\circ}_{}$ and $\sigma=167.2^{\circ}_{}$. Meanwhile, the model predictions for the effective neutrino masses $m^{}_{\beta}$ and $m^{}_{\beta\beta}$ are found to be $9.020~{\rm meV}$ and $2.282~{\rm meV}$, respectively.
	
	Furthermore, a brief illustration on why the IO case is excluded in \emph{Case~A} is helpful. In fact, the observed values of $\Delta m^2_{21}$ and $\Delta m^2_{32}$ impose strong constraints on the parameter space of $\tau$ in the IO case. To be specific, if both $\Delta m^2_{21}$ and $\Delta m^2_{32}$ are within their individual $3\sigma$ ranges from global-fit results, the allowed range of $\{{\rm Re}\,\tau,{\rm Im}\,\tau\}$ is restricted to be in a small ring centered on $\tau = 1/2+{\rm i}\,\sqrt{3}/2$, where the predicted values of $\theta^{}_{13}$ are found to be around either 0 or $90^{\circ}_{}$. Therefore the IO case is not compatible with the neutrino oscillation data.

\begin{figure}[t!]
	\centering		\includegraphics[width=0.98\textwidth]{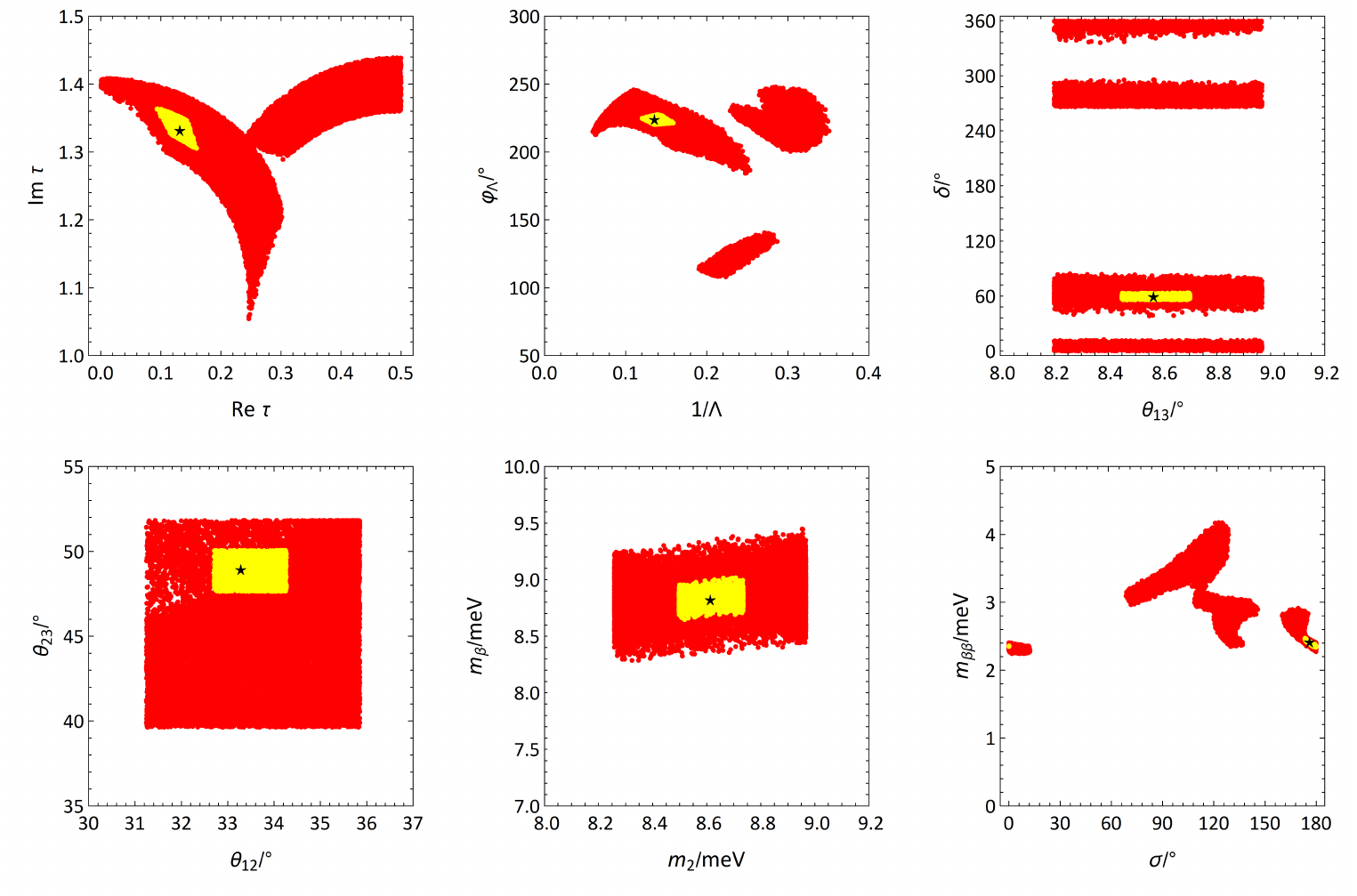}
	\vspace{0cm}
	\caption{The allowed parameter space of model parameters, as well as the constraints on the low-energy observables in \emph{Case~B} with $\widetilde{\gamma}=0$, where the $1\sigma$ (yellow dots) and $3\sigma$ (red dots) ranges of flavor mixing angles and neutrino mass-squared differences from the global-fit analysis of neutrino oscillation data have been input~\cite{Esteban:2020cvm}. The best-fit values from the $\chi^2_{}$-fit analysis are denoted by the black stars. }
	\label{fig:MSM_R2} 
\end{figure}

\item In \emph{Case~B} with $\widetilde{\gamma} = 0$, the total number of model parameters is also reduced to be eight. Unlike \emph{Case~A},  we find that \emph{Case~B} is compatible with neutrino oscillation data even at the $1\sigma$ level in the NO case. The allowed ranges of model parameters as well as the constrained ranges of low-energy observables are shown in Fig.~\ref{fig:MSM_R2}. As one can see from the top-left panel, there are two separated parts in the $3\sigma$ allowed parameter space of $\{{\rm Re}\,\tau, {\rm Im}\,\tau\}$, corresponding to two distinct allowed hierarchies with $b^{}_1>b^{}_2>1$ and $b^{}_1>1>b^{}_2$ in the charged-lepton sector, respectively. While at the $1\sigma$ level, only one narrow region remains in the allowed parameter space of  $\{{\rm Re}\,\tau, {\rm Im}\,\tau\}$, where $0.1 \lesssim {\rm Re}\,\tau \lesssim 0.16$ and $1.31 \lesssim {\rm Im}\,\tau \lesssim 1.36$. Similarly, there appear three separated regions in the $3\sigma$ allowed parameter space of $\{\Lambda^{-1}, \phi^{}_{\Lambda}\}$ in the top-middle panel of Fig.~\ref{fig:MSM_R2}. However, as has been mentioned before, under the transformations ${\rm Re}\,\tau \rightarrow -{\rm Re}\,\tau$ and $\phi^{}_{\Lambda} \rightarrow 2\pi-\phi^{}_{\Lambda}$, the predictions for low-energy observables keep unchanged, except that the signs of all CP-violating phases are reversed. Therefore the rightmost two parts in the top-middle panel seem to be connected to each other by the transformation $\phi^{}_{\Lambda} \rightarrow 2\pi-\phi^{}_{\Lambda}$. In the top-right panel, the allowed values of $\delta$ are lying in several separated regions, probably originating from different regions in the allowed parameter space of $\{\Lambda^{-1}_{},\varphi^{}_{\Lambda}\}$. In \emph{Case~B}, the predicted values of three mixing angles can essentially saturate their individual $3\sigma$ ranges allowed by neutrino oscillation data, and there is no significant correlation among them. The $3\sigma$ allowed ranges of $m^{}_{\beta}$ and $m^{}_{\beta\beta}$ are larger than those in \emph{Case~A}, which is mainly due to the fact that the allowed range of $\theta^{}_{12}$ becomes larger in \emph{Case~B}. Performing the $\chi^2_{}$-fit analysis in \emph{Case~B}, we find that the minimum $\chi^2_{\rm min}=0.0741$ is achieved at
\begin{eqnarray}
{\rm Re}\,\tau = 0.1320\;, \quad {\rm Im}\,\tau = 1.331 \;, \quad \Lambda \ = 7.362 \;, \quad \varphi^{}_{\Lambda} = 223.4^{\circ}_{} \;,
\label{eq:bfcaseB}
\end{eqnarray}
which together with the charged-lepton masses lead to $\mu^{}_l = 1.853 \times 10^{-4}~{\rm GeV}$, $b^{}_1 = 1.872 \times 10^{3}$, $b^{}_2 = 64.30$ and $\mu^{}_{\nu}=29.47~{\rm meV}$. Furthermore, we get the neutrino mass spectrum $m^{}_{1} = 0$, $m^{}_{2}=8.625~{\rm meV}$ and $m^{}_{3}=50.18~{\rm meV}$, three mixing angles $\theta^{}_{12}=33.28^{\circ}_{}$, $\theta^{}_{13}=8.567^{\circ}_{}$, $\theta^{}_{23} = 48.88^{\circ}_{}$ and two CP-violating phases $\delta=58.67^{\circ}_{}$ and $\sigma=176.1^{\circ}_{}$. Meanwhile, the predictions for two effective neutrino masses $m^{}_{\beta}$ and $m^{}_{\beta\beta}$ are $8.827~{\rm meV}$ and $2.404~{\rm meV}$, respectively. The remarkable difference between {\it Case A} and {\it Case B} is their predictions for the octant of $\theta^{}_{23}$. \emph{Case~B} shows no clear preference for the octant of $\theta^{}_{23}$, while \emph{Case~A} prefers the first octant. If the present hint for the second octant of $\theta^{}_{23}$ is confirmed by future neutrino oscillation data, {\it Case A} will be definitely ruled out. In addition, the precision measurement of $\theta^{}_{12}$ and the determination of the CP-violating phase $\delta$ will also shed some light on the discrimination between these two simple but viable scenarios.
\end{itemize}

In summary, by a detailed numerical analysis, we have demonstrated that our model with ten real parameters is well compatible with current neutrino oscillation data. Even more, if the complex parameter $\widetilde{\Lambda}$ in {\it Case A} or $\widetilde{\gamma}$ in {\it Case B} is set to zero, the model is still allowed by experimental observations.

\subsection{Analytical results}

It is worthwhile to notice that there are no additional free parameters other than the complex modulus $\tau$ in the neutrino sector in \emph {Case~A} with $\widetilde{\Lambda} = 0$. As a result, the effective Majorana neutrino mass matrix $M^{}_\nu$ in this case turns out to be simple enough, rendering analytical calculations under some reasonable approximations to be possible. Analytical calculations, though approximate, will be helpful for understanding the flavor structures of lepton mass matrices.

First, as we have seen from the numerical calculations in the previous section, the allowed values of $\rm{Im}\, \tau$ are located in a very narrow region $1.326 \lesssim \rm{Im}\, \tau \lesssim 1.352$, for which $|q| = e^{-2\pi {\rm Im}\, \tau}$ is small enough and thus it is safe to retain only the leading-order terms in the $q$-expansions of all the modular forms. For this purpose, we introduce two auxiliary real parameters
\begin{eqnarray}
x\equiv {\rm exp}\left(-\frac{2\pi}{5}\; {\rm Im}\,\tau\right) \; , \quad y \equiv \frac{2\pi}{5}\; \rm{Re}\,\tau \; ,
\label{eq:xy}
\end{eqnarray}
which are actually the modulus and argument of $q^{1/5}$ (i.e., $q^{1/5} \equiv x e^{{\rm i}y}$), and six basis vectors given in Eq.~(\ref{eq:basisq}) approximate to
\begin{eqnarray}
\widehat{e}^{}_1 \approx 1 \; , \quad
\widehat{e}^{}_2 \approx x e^{{\rm i}y}_{} \; , \quad
\widehat{e}^{}_3 \approx x^2_{} e^{2{\rm i} y}_{} \; , \quad
\widehat{e}^{}_4 \approx x^3_{} e^{3{\rm i} y}_{} \; , \quad
\widehat{e}^{}_5 \approx x^4_{} e^{4{\rm i} y}_{} \; , \quad
\widehat{e}^{}_6 \approx x^5_{} e^{5{\rm i} y}_{} \; , \quad
\label{eq:eiapprox}
\end{eqnarray}
where $x$ serves as the expansion parameter. To have a ballpark feeling about the size of $x$, we take the best-fit value of ${\rm Im}\, \tau = 1.341$, as found in Eq.~(\ref{eq:bfcaseA}), and then obtain $x \approx 0.185$ from Eq.~(\ref{eq:xy}), which is not perfect but reasonably good for perturbation calculations.

Then, applying the approximate expressions in Eq.~(\ref{eq:eiapprox}) to $M_{\rm D}^{}$ and $M_{\rm R}^{}$ in Eq.~(\ref{eq:matrices}), one can get the analytical form of the effective neutrino mass matrix via $M_{\nu}^{} = - M_{\rm D}^{}M_{\rm R}^{-1} M_{\rm D}^{\rm T}$, i.e.,
\begin{eqnarray}
M_{\nu}^{} \approx \mu^{}_{\nu} \left[ \frac{5\sqrt{3}}{96} \left(\begin{matrix} 2 & 0 & 0 \cr 0 & 0 & -1 \cr 0 & -1 & 0 \end{matrix}\right) +
\frac{\sqrt{3}e^{3{\rm i}y}}{480\,x^3} \left(
\begin{matrix}
0 & - 5\sqrt{2}\, x^2 e^{-2{\rm i}y}  & 80\sqrt{2}\, x^4 e^{-4{\rm i} y}  \cr - 5\sqrt{2}\, x^2 e^{-2{\rm i}y} & x e^{-{\rm i}y} & 0 \cr 80\sqrt{2}\, x^4 e^{-4{\rm i} y} & 0 & 3 \\
\end{matrix}
\right) \right] \; , \quad
\label{eq:Mnuapp}
\end{eqnarray}
where $\mu^{}_\nu$ denotes the absolute scale of neutrino masses and only the terms up to ${\cal O}(x^4)$ in the second matrix in the square parentheses are kept. Some comments on the general structure of $M^{}_\nu$ are helpful. In Eq.~(\ref{eq:Mnuapp}), we have intentionally divided the neutrino mass matrix into two parts, the first of which is constant. Considering the absolute value of the ratio of the coefficient in front of the first matrix on the right-hand side of Eq.~(\ref{eq:Mnuapp}) to that of the second matrix, namely, $25\, x^3 \approx 0.158$, we can see that both of these two parts make remarkable contributions to $M^{}_\nu$. However, one can observe from Eq.~(\ref{eq:Mnuapp}) that only the $(3,3)$-element of $M^{}_\nu$ survives at the leading order, whose absolute value is approximately the largest mass eigenvalue $m^{}_3$ in the NO case. On the other hand, the lightest neutrino must be massless in the MSM, i.e., $m^{}_1 = 0$. Therefore, it is straightforward to derive three neutrino mass eigenvalues
\begin{eqnarray}
m^{}_1 = 0 \; , \quad m^{}_2 \approx \left( 1 + 50\, x^2 \right) \frac{\sqrt{3}}{480\,x^2} \mu_{\nu}^{} \; , \quad m^{}_3 \approx \frac{\sqrt{3}}{160\,x^3}\mu_{\nu}^{} \; ,
\label{eq:anami}
\end{eqnarray}
from which one can determine the neutrino mass ratio
\begin{eqnarray}
r \equiv \frac{m^{}_2}{m^{}_3} \approx \frac{x}{3} \left(1 + 50\, x^2\right) \; .
\label{eq:rana}
\end{eqnarray}
When setting $x \approx 0.185$ from the best-fit value of ${\rm Im} \, \tau=1.341$, we find that the prediction from Eq.~(\ref{eq:rana}) is $r \approx 0.167$, which is in excellent agreement with the best-fit value $r \approx 0.165$.

As the symmetric and complex neutrino mass matrix is diagonalized by the unitary matrix $U^{}_\nu$ via $U_{\nu}^{\dagger} M_{\nu}^{} U_{\nu}^{*} = {\rm Diag}\{m^{}_1, m^{}_2, m^{}_3\}$, we can also obtain
\begin{eqnarray}
U_{\nu}^{} \approx \left(\begin{matrix} e^{-3{\rm i}y/2} & 0 & 0 \cr 0 & e^{-{\rm i}y/2} & 0 \cr 0 & 0 & 1\end{matrix}\right) \cdot \left(
\begin{matrix}
\cos\theta^{}_\nu & - \sin\theta^{}_\nu & 0 \cr
+ \sin\theta^{}_\nu & \cos\theta^{}_\nu & 0 \cr
0 & 0 & 1
\end{matrix}
\right) \cdot \left( \begin{matrix} e^{-5{\rm i}y/2} & 0 & 0 \cr 0 & 1 & 0 \cr 0 & 0 & 1 \end{matrix} \right)\; ,
\label{eq:mixana}
\end{eqnarray}
where the rotation angle $\theta^{}_\nu$ is determined by $\tan\theta^{}_\nu = 5\sqrt{2}\, x$. Given $x \approx 0.185$, one can estimate $\theta^{}_\nu = \arctan(5\sqrt{2}\, x) \approx 52.60^\circ$, which is quite large. Furthermore, with the help of Eq.~(\ref{eq:rana}), we can immediately establish a rather simple relation between the rotation angle and the neutrino mass ratio $m^{}_2/m^{}_3 \approx \tan\theta^{}_\nu \sec^2\theta^{}_\nu /(15\sqrt{2})$. We have numerically checked that the approximate analytical result of $U^{}_\nu$ in Eq.~(\ref{eq:mixana}) agrees very well with the exact result, and the difference arises from the omission of higher-order terms.

Next, we proceed with the charged-lepton sector. Up to the order of $\mathcal{O}(x^3)$, the charged-lepton mass matrix $M^{}_l$ in Eq.~(\ref{eq:matrices}) can be expressed as
\begin{eqnarray}
M^{}_l \approx \mu^{}_{l}
\left(
\begin{matrix}
-3 & -5\sqrt{3} & 3\sqrt{2}\gamma e_{}^{{-\rm i}\varphi^{}_{\gamma}} \cr
0 & 0 & 0 \cr
-15\sqrt{2}\, x e_{}^{{-\rm i}y} & 5\sqrt{6}\, x e_{}^{{-\rm i}y} & 30\left(3 + \gamma e_{}^{{-\rm i} \varphi^{}_{\gamma}}\right) x e_{}^{{-\rm i}y}
\end{matrix}
\right) \cdot
\left(
\begin{matrix}
b^{}_1 & 0 & 0\\
0 & b^{}_2 & 0 \\
0 & 0 & 1
\end{matrix}
\right) \; ,
\label{eq:mlapp}
\end{eqnarray}
while the corresponding Hermitian matrix $H^{}_l \equiv M^{}_l M^\dagger_l$ reads
\begin{eqnarray}
H^{}_l \approx \mu^2_l \left( \begin{matrix} 75b^2_2 + 18\gamma^2 & 0 & 15\sqrt{2}\left[6\gamma(\gamma + 3e^{-{\rm i}\varphi^{}_\gamma}) - 5 b^2_2\right] x e^{{\rm i}y} \cr 0 & 0 & 0 \cr 15\sqrt{2}\left[6\gamma(\gamma + 3e^{{\rm i}\varphi^{}_\gamma}) - 5 b^2_2\right]x e^{-{\rm i}y}  & 0 & 150 x^2 \left[b^2_2 + 6(\gamma^2 + 6\gamma \cos\varphi^{}_\gamma + 9)\right] \end{matrix} \right) \; ,
\label{eq:hlapp}
\end{eqnarray}
where $b^2_1$ has been set to be zero because of the strong hierarchy $b^2_1 \ll b^2_2 \ll 1$ as indicated by numerical calculations in the previous section. If $b^2_2 = 0$ is further assumed, then one can verify that the masses of the first two generations of charged leptons are vanishing, i.e., $m^{}_e = m^{}_\mu = 0$. Therefore, the terms associated with $b^2_2$ will be retained. It is worth stressing that even if $b^2_1$ is set to be nonzero, the electron mass is still vanishing due to the special structure of $M^{}_l$ in Eq.~(\ref{eq:mlapp}). To generate a nonzero electron mass, we have to keep higher-order terms of $x$ in $M^{}_l$. For illustration, we work with the accuracy of ${\cal O}(x^3)$ and thus focus on the charged-lepton mass matrix $M^{}_l$ in Eq.~(\ref{eq:mlapp}) and accordingly $H^{}_l$ in Eq.~(\ref{eq:hlapp}), implying $m^{}_e = 0$. As the Hermitian matrix $H^{}_l$ is diagonalized by the unitary matrix $U^{}_l$ via  $U_{l}^{\dagger} H_{l}^{} U_{l}^{} = {\rm Diag}\{m_e^2, m_{\mu}^2, m_{\tau}^2\}$, we can obtain
\begin{eqnarray}
U_{l}^{} =
\left(
\begin{matrix}
0 & - \sin\theta^{}_l e^{{\rm i}(y - \xi)} &  \cos \theta^{}_l \cr
1 & 0 & 0 \cr
0 & \cos\theta^{}_l & \sin\theta^{}_l e^{-{\rm i}(y - \xi)}
\end{matrix}
\right) \; ,
\label{eq:ulapp}
\end{eqnarray}
where the rotation angle $\theta^{}_l$ is determined by $\tan^2 \theta^{}_l = 50\, x^2 (\gamma^2 + 6\gamma \cos\varphi^{}_\gamma + 9)/\gamma^2$ and the phase by $\tan\xi = 3 \sin\varphi^{}_\gamma/( 3\cos\varphi^{}_\gamma+\gamma)$. Given the best-fit values of model parameters, we have $\theta_{l}^{}\approx 48.17^{\circ}$ and $\xi\approx 23.16^{\circ}$. Meanwhile, three charged-lepton mass eigenvalues are found to be $m^{}_e = 0$ and
\begin{eqnarray}
m^2_\mu \approx 75\, b_2^2\, x_{}^2 \gamma^{-4} \left(25  +
   20 \gamma \cos\varphi_{\gamma}^{}+4 \gamma_{}^2\right) m^2_\tau\left|\cos 2\theta^{}_l\right|\; , \quad m_{\tau}^2 \approx 18 \gamma_{}^2 \mu^2_l \sec^2\theta^{}_l \; ,
\label{eq:mutauapp}
\end{eqnarray}
where the rotation angle $\theta^{}_l$ has been defined below Eq.~(\ref{eq:ulapp}). Substituting the best-fit values of relevant parameters in Eq.~(\ref{eq:mutauapp}), one obtains $m^{}_e = 0$, $m^{}_{\mu}/\mu^{}_{l} = 2.044$ and $m^{}_{\tau}/\mu^{}_{l} = 47.862$, while the exact numerical results are $(m^{}_e, m^{}_{\mu}, m^{}_{\tau})/\mu^{}_{l} = (0.013, 2.764, 48.924)$. Bearing in mind that only the leading-order terms are kept in the charged-lepton mass matrix, we can see a reasonably good agreement between analytical and numerical results.

Finally, with the unitary matrices $U^{}_\nu$ in Eq.~(\ref{eq:mixana}) and $U^{}_l$ in Eq.~(\ref{eq:ulapp}), the PMNS matrix is thus given by
\begin{eqnarray}
U = U^\dagger_l U^{}_\nu \approx \left( \begin{matrix} \sin\theta^{}_\nu & \cos\theta^{}_\nu & 0 \cr -\sin\theta^{}_l \cos\theta^{}_\nu & \sin\theta^{}_l \sin\theta^{}_\nu & \cos\theta^{}_l \cr \cos\theta^{}_l \cos\theta^{}_\nu & - \cos\theta^{}_l \sin\theta^{}_\nu & \sin\theta^{}_l \end{matrix} \right) \cdot \left( \begin{matrix} 1 & 0 & 0 \cr 0 & e^{-{\rm i}(5y/2 - \xi)} & 0 \cr 0 & 0 & 1\end{matrix}\right) \; ,
\label{eq:PMNSapp}
\end{eqnarray}
where the unphysical phases have been eliminated by redefining the charged-lepton fields and the neutrino field with mass $m^{}_1 = 0$. From the PMNS matrix in Eq.~(\ref{eq:PMNSapp}), we can extract three flavor mixing angles as below
\begin{eqnarray}
\sin\theta^{}_{13} \approx 0 \; , \quad  \sin\theta^{}_{12} \approx \cos\theta^{}_\nu \; , \quad \sin \theta^{}_{23} \approx \cos\theta^{}_l \; ,
\label{eq:mixingapp}
\end{eqnarray}
and the Majorana CP phase $\sigma \approx \xi - 5y/2$. Interestingly, the rotation angle $\theta^{}_\nu$ from the neutrino sector and $\theta^{}_l$ from the charged-lepton sector are directly related to the mixing angle $\theta^{}_{12}$ and $\theta^{}_{23}$ via the simple relation $\theta^{}_{12} \approx \pi/2 - \theta^{}_\nu$ and $\theta^{}_{23}  \approx \pi/2 - \theta^{}_l$, respectively. Substituting the best-fit values of model parameters shown in Eq.~(\ref{eq:bfcaseA}), we find that our approximate analytical results lead to
\begin{eqnarray}
\theta_{13} \approx 0\;,\quad
\theta_{12} \approx 37.40^{\circ}\;,\quad
\theta_{23} \approx 41.83^{\circ}\;, \quad
\sigma \approx 165.8^{\circ} \; ,
\label{eq:bfapp}
\end{eqnarray}
where one can see that the values of $\theta^{}_{12}$, $\theta^{}_{23}$ and $\sigma$ are in good agreement with their individual best-fit values shown in the paragraph below Eq.~(\ref{eq:bfcaseA}). In addition, recalling approximate analytical expression of $r$ and $m^{}_{\tau}$ in Eqs.~(\ref{eq:rana}) and (\ref{eq:mutauapp}), we can see that the free model parameters $x$ and $\gamma\mu_{l}^{}$ are directly related to the observables by
\begin{eqnarray}
x=3 \sqrt{\frac{\Delta m^{2}_{21}}{\Delta m^{2}_{31}}}\sin_{}^2\theta_{12}^{} \; ,\quad
\gamma\mu_{l}^{}=\frac{m_{\tau}\sin\theta_{23}^{}}{3\sqrt{2}} \; .
\label{eq:theta_mass}
\end{eqnarray}

Before closing this subsection, let us briefly discuss how to generate a nonzero $\theta_{13}$. Note that the unitary matrix $U_{\nu}$ in the neutrino sector in Eq.~(\ref{eq:mixana}) is only kept to $\mathcal{O}(x)$ and thus parametrized by a pure $(1,2)$-rotation. If we improve our calculations with the accuracy of $\mathcal{O}(x^3)$, $U_{\nu}$ will be modified with an extra $(2,3)$-rotation, which can be expressed as
\begin{eqnarray}
U_{\nu}^{} \approx
\left(\begin{matrix} e^{-4{\rm i}y} & 0 & 0 \cr 0 & e^{-3{\rm i}y} & 0 \cr 0 & 0 & 1\end{matrix}\right) \cdot
\left(
\begin{matrix}
1 & 0 & 0 \cr
0 & \cos\theta^{\prime}_{\nu} & -\sin\theta^{\prime}_{\nu}\cr
0 & + \sin\theta^{\prime}_{\nu}& \cos\theta^{\prime}_{\nu}
\end{matrix}
\right)\cdot
\left(
\begin{matrix}
\cos\theta^{}_\nu & - \sin\theta^{}_\nu & 0 \cr
+ \sin\theta^{}_\nu & \cos\theta^{}_\nu & 0 \cr
0 & 0 & 1
\end{matrix}
\right) \cdot \left( \begin{matrix} 1 & 0 & 0 \cr 0 & e^{5{\rm i}y/2} & 0 \cr 0 & 0 & 1 \end{matrix} \right)\;,
\end{eqnarray}
where the rotation angle $\theta^{\prime}_{\nu}$ is determined by $\sin\theta_{\nu}^{\prime} = 25x^3_{}/3$. Adopting the best-fit value of $x\approx 0.185$, we obtain $\theta_{\nu}^{\prime} =\arcsin(25x^3_{}/3)\approx 3.023^\circ_{}$, indicating $\theta^{\prime}_{\nu}$  is quite a small angle. Then the modified version of the PMNS matrix reads
\begin{eqnarray}
U=U_{l}^{\dagger}U_{\nu}^{}
\approx
\left( \begin{matrix} \sin\theta^{}_\nu & \cos\theta^{}_\nu & \sin\theta_{\nu}^{\prime}e^{-{\rm i}(5y-\xi+\pi)} \cr -\sin\theta^{}_l \cos\theta^{}_\nu & \sin\theta^{}_l \sin\theta^{}_\nu & \cos\theta^{}_l \cr \cos\theta^{}_l \cos\theta^{}_\nu & - \cos\theta^{}_l \sin\theta^{}_\nu & \sin\theta^{}_l \end{matrix} \right) \cdot \left( \begin{matrix} 1 & 0 & 0 \cr 0 & e^{-{\rm i}(5y/2 - \xi)} & 0 \cr 0 & 0 & 1\end{matrix}\right) \;,
\label{eq:mod PMNS}
\end{eqnarray}
where one can observe that the $(1,3)$-element of $U$ acquires a small nonzero value proportional to $\sin\theta^{\prime}_\nu$ while all the other elements remain to be approximately unchanged. Thanks to this small correction, we can obtain the approximate expressions of $\theta^{}_{13}$ and $\delta$ as
\begin{eqnarray}
\theta^{}_{13} \approx \theta_{\nu}^{\prime}\;,\quad
\delta \approx 5y-\xi + \pi\;.
\label{eq:theta13app}
\end{eqnarray}
Substituting the best-fit values of model parameters shown in Eq.~(\ref{eq:bfcaseA}) into Eq.~(\ref{eq:theta13app}), one can arrive at
\begin{eqnarray}
\theta_{13}\approx 3.023^{\circ}\;,\quad
\delta\approx 231.6^{\circ}\;,
\label{eq:theta13deltanum}
\end{eqnarray}
where the value of $\theta^{}_{13}$ is still much smaller than its best-fit value. However, if we retain the terms of $\mathcal{O}(x^4)$ in $M_l^{}$ in Eq.~(\ref{eq:mlapp}) from the very beginning, then a nonzero value of $m_e^{}$ will be obtained and $\theta_{13}$ will receive an additional correction from the charged-lepton sector. In fact, we have checked that $\theta_{13}\approx 9.012^{\circ}$ can indeed be generated at this order of approximations if the best-fit values of model parameters are taken.

\section{Summary}\label{sec:sum}

The double covering of modular groups can accommodate the modular forms with odd weights, and thus provide us with new possibilities to account for tiny neutrino masses, lepton flavor mixing and CP violation. In this paper, we investigate the basic properties of the double covering of modular $\Gamma^{}_5 \simeq A^{}_{5}$ group, i.e., the modular $A^{\prime}_{5}$ group, which has not been studied in the previous literature. As a practical application, we have considered the minimal seesaw model with a modular $A^{\prime}_{5}$ symmetry, in which we numerically explore the allowed parameter space and analytically study the mass spectrum and flavor mixing in the lepton sector with some reasonable assumptions.

The main results are summarized as follows. First of all, we begin with the modular forms of weight one. With the help of the Dedekind eta function and Klein form, we obtain the basis vectors of the modular space ${\cal M}^{}_1 [\Gamma(5)]$, and the modular forms $Y^{(1)}_{\widehat{\bf 6}}$ with weight one turn out to be the linear combination of these basis vectors. Next we  derive the modular forms of weights up to six using the Kronecker product rules of $A^{\prime}_5$ and present their explicit expressions. After the modular forms are determined, we proceed to apply the double-covering group $A^\prime_5$ to the concrete models. There exists a two-dimensional irreducible representation $\widehat{\bf 2}^{\prime}_{}$ in the group $A^{\prime}_{5}$, into which we assign two right-handed neutrino singlets $\widehat{N}^{\rm C}_{}$ in the minimal seesaw model, and the charged-lepton doublets $\widehat{L}$ and three charged-lepton singlets $ \{ \widehat{E}^{\rm C}_{1}, \widehat{E}^{\rm C}_{2}, \widehat{E}^{\rm C}_{3} \}$ are assumed to transform as {\bf 3} and three one-dimensional representations of $A^{\prime}_{5}$, respectively. In the most general case, the model contains ten real parameters, which are the modulus parameter $\tau = {\rm Re}\,\tau+{\rm i}\, {\rm Im}\,\tau$ together with the parameters $\mu^{}_l \equiv v^{}_{\rm d}\gamma^{}_3/\sqrt{2} $, $b^{}_1 \equiv \gamma^{}_1/\gamma^{}_3$, $b^{}_2 \equiv \gamma^{}_2/\gamma^{}_3$ and $\widetilde{\gamma} \equiv \gamma e^{{\rm i}\varphi^{}_{\gamma}}$ in the charged-lepton sector and $\mu^{}_{\nu} \equiv g^{2}_{}v^{2}_{\rm u}/(2\Lambda^{}_{1})$ and $\widetilde{\Lambda} \equiv \Lambda e^{{\rm i}\varphi^{}_{\Lambda}}$ in the neutrino sector. We find that our model is consistent with the global-fit results of neutrino oscillation data at the $1\sigma$ level only in the NO case. The best-fit values, together with $1\sigma$ and $3\sigma$ allowed ranges of model parameters and low-energy observables, are also given.

In addition, we have investigated two simple but viable cases, which are \emph{Case A} with $\widetilde{\Lambda}=0$ and \emph{Case B} with $\widetilde{\gamma}=0$. In these two cases, only eight real model parameters are involved. Numerically we find that \emph{Case A} is compatible with the oscillation data at the $3\sigma$ level in the NO case while \emph{Case B} can be consistent with the global-fit results within the $1\sigma$ level in the NO case. In particular, the effective Majorana neutrino mass matrix $M^{}_{\nu}$ in \emph{Case A} turns out to be phenomenologically appealing, since no additional parameters other than the modulus $\tau$ are present. This allows us to perform analytical calculations under some reasonable approximations. Expanding lepton mass matrices in terms of the parameter $x \equiv {\rm exp}[-(2\pi\, {\rm Im}\,\tau)/5]$, which is about 0.185 given the best-fit value of ${\rm Im}\,\tau \approx 1.341$, we show that the PMNS matrix up to the order of ${\cal O}(x^2_{})$ can be described by the combination of two rotations coming from the neutrino sector with the rotation angle $\theta^{}_{\nu} = \arctan(5\sqrt{2}x) \approx 52.60^\circ$ and the charged-lepton sector with the angle $\theta^{}_l = \arctan[5\sqrt{2}x \sqrt{\gamma^2_{}+6\gamma \cos\varphi^{}_{\gamma}+9}/\gamma] \approx 48.17^\circ$, respectively. As a consequence, we obtain simple expressions of the mixing angles $\theta^{}_{12}$ and $\theta^{}_{23}$, namely, $\theta^{}_{12} \approx \pi/2 - \theta^{}_{\nu}$ and $\theta^{}_{23} \approx \pi/2 -\theta^{}_l$, which agree well with their individual numerical results. A nonzero $\theta^{}_{13}$ can be generated only if the higher-order corrections are taken into account. At the order of ${\cal O}(x^3_{})$, an additional rotation with the rotation angle $\theta^{\prime}_{\nu}=\arcsin(25x^3_{}/3)$ in the neutrino sector will contribute to the PMNS matrix, leading to $\theta^{}_{13} \approx \theta^{\prime}_{\nu} \approx 3.023^\circ$. Furthermore, the approximate expressions of two CP-violating phases $\delta$ and $\sigma$ have also been gained.

For further exploration along this direction, it will be interesting to bring the modular $A^{\prime}_{5}$ group into the model building of both lepton and quark masses, and give a unified description of both quark and lepton masses, flavor mixing patterns and CP violation. We hope to come back to this possibility in the near future.

\section*{Acknowledgements}

This work was supported in part by the National Natural Science Foundation of China under grant No.~11775232 and No.~11835013, and by the CAS Center for Excellence in Particle Physics.

\newpage
\appendix
\section{Conjugacy classes and representation matrices}\label{app:A}

\renewcommand\arraystretch{1.2}
\begin{table}[t]
	\begin{center}
		\vspace{-0.25cm} \caption{The character table of the group $A^{\prime}_{5}$.} \vspace{0.5cm}
		\begin{tabular}{c|c|c|c|c|c|c|c|c|c}
			\hline\hline $A^{\prime}_{5}$ & {\bf 1} & {\bf 3} & ${\bf 3^{\prime}_{}}$ & {\bf 4} & {\bf 5} & $\widehat{\bf 2}$ & $\widehat{\bf 2^{\prime}_{}}$ & $\widehat{\bf 4}$ & $\widehat{\bf 6}$ \\
			\hline $1C^{}_1$ & 1 & 3 & 3 & 4 & 5 & 2 & 2 & 4 & 6 \\
			 $12 C^{}_{5}$ & 1 & $\phi$ & $1-\phi$ & $-1$ & 0 & $-\phi$ & $\phi-1$ & $-1$ & $1$ \\
			 $12 C_{5}^{\prime}$ & 1 & $1-\phi$ & $\phi$ & $-1$ & 0 & $\phi-1$ & $-\phi$ & $-1$ & 1 \\
			$20 C^{}_{3}$ & 1 & 0 & 0 & 1 & $-1$ & $-1$ & $-1$ & $1$ & 0 \\
			$30 C^{}_{4}$ & 1 & $-1$ & $-1$ & 0 & 1 & 0 & 0 & 0 & 0 \\
			\hline $1C^{}_2$ & 1 & 3 & 3 & 4 & 5 & $-2$ & $-2$ & $-4$ & $-6$ \\
			$12 C^{}_{10}$ & 1 & $\phi$ & $1-\phi$ & $-1$ & 0 & $\phi$ & $1-\phi$ & $1$ & $-1$ \\
			$12 C_{10}^{\prime}$ & 1 & $1-\phi$ & $\phi$ & $-1$ & 0 & $1-\phi$ & $\phi$ & 1 & $-1$ \\
			$20 C^{}_{6}$ & 1 & 0 & 0 & 1 & $-1$ & $1$ & $1$ & $-1$ & 0 \\
			\hline\hline
		\end{tabular}
		\label{table:character_table}
	\end{center}
\end{table}
\renewcommand\arraystretch{1}
As has been mentioned in Sec.~\ref{sec:double}, the group $A^\prime_5$ has 120 elements, which can be divided into the following nine conjugacy classes~\cite{Hashimoto:2011tn}
\begin{eqnarray}
1 C^{}_{1}&:& 1; \nonumber \\
12 C^{}_{5}&:& T, T^{4}_{}, S T^{2}_{}R, T^{2}_{} S R, S T^{3}_{}, T^{3}_{} S, S T S R, T S T R, T S T^{2}_{}, T^{2}_{} S T, T^{3}_{} S T^{4}_{} R, T^{4}_{} S T^{3}_{} R; \nonumber \\
12 C_{5}^{\prime}&:& T^{2}_{}, T^{3}_{}, S T^{2}_{} S R, S T^{3}_{} S R,(S T^{2}_{})^{2}_{},(T^{2}_{} S)^{2}_{},(S T^{3}_{})^{2}_{},(T^{3}_{} S)^{2}_{},(T^{2}_{} S)^{2}_{} T^{3}_{}R, T^{3}_{}(S T^{2}_{})^{2}_{}R,  \nonumber \\
&& T^{3}_{} S T^{2}_{} S T^{4}_{}, T^{4}_{} S T^{2}_{} S T^{3}_{}; \nonumber \\
20 C^{}_{3}&:& S T, T S, S T^{4}_{} R, T^{4}_{} S R, T S T^{3}_{} R, T^{2}_{} S T^{2}_{} R, T^{2}_{} S T^{4}_{}, T^{3}_{} S T R, T^{3}_{} S T^{3}_{}, T^{4}_{} S T^{2}_{}, TST^3_{}S R,\nonumber \\
&&  T^2_{} S T^{3}_{} S, T^{3}_{} S T^{2}_{} S, S T^{2}_{} S T^{3}_{}, S T^{3}_{} S T R, S T^{3}_{} S T^{2}_{},(T^{2}_{} S)^{2}_{} T^{2}_{}R, T^{2}_{}(T^{2}_{} S)^{2}_{}R,(S T^{2}_{})^{2}_{} S, \nonumber \\
&& (S T^{2}_{})^{2}_{} T^{2}_{} R; \nonumber \\
30 C^{}_{4}&:& S T^{2}_{} S T^{3}_{} S, T S T^{4}_{}, T^{4}_{}(S T^{2}_{})^{2}_{}, T^{2}_{} S T^{3}_{},(T^{2}_{} S)^{2}_{} T^{3}_{} S, S T^{2}_{} S T, S, T^{3}_{} S T^{2}_{} S T^{3}_{}, \nonumber \\
&& T^{3}_{} S T^{2}_{} S T^{3}_{} S, T^{3}_{} S T^{2}_{}, T^{4}_{} S T^{2}_{} S T^{3}_{} S, T S T^{2}_{} S, S T^{3}_{} S T^{2}_{} S, T^{4}_{} S T,(T^{2}_{} S)^{2}_{} T^{4}_{}, \nonumber \\
&& S T^{2}_{} S T^{3}_{} S R, T S T^{4}_{} R , T^{4}_{}(S T^{2}_{})^{2}_{}R, T^{2}_{} S T^{3}_{}R,(T^{2}_{} S)^{2}_{} T^{3}_{} SR, S T^{2}_{} S TR, SR,  \nonumber \\
&& T^{3}_{} S T^{2}_{} S T^{3}_{}R, T^{3}_{} S T^{2}_{} S T^{3}_{} SR, T^{3}_{} S T^{2}_{}R, T^{4}_{} S T^{2}_{} S T^{3}_{} SR, T S T^{2}_{} SR, S T^{3}_{} S T^{2}_{} SR,  \nonumber \\
&& T^{4}_{} S TR,(T^{2}_{} S)^{2}_{} T^{4}_{}R ; \nonumber \\
1C^{}_2 &:& R; \nonumber \\
12 C^{}_{10} &:& TR, T^{4}_{}R, S T^{2}_{}, T^{2}_{} S, S T^{3}_{}R, T^{3}_{} SR, S T S, T S T, T S T^{2}_{}R, T^{2}_{} S TR,  T^{3}_{} S T^{4}_{}, T^{4}_{} S T^{3}_{};\nonumber \\
12 C_{10}^{\prime} &:& T^{2}_{}R, T^{3}_{}R, S T^{2}_{} S, S T^{3}_{} S,(S T^{2}_{})^{2}_{}R,(T^{2}_{} S)^{2}_{}R,(S T^{3}_{})^{2}_{}R,(T^{3}_{} S)^{2}_{}R, (T^{2}_{} S)^{2}_{} T^{3}_{},  \nonumber \\
&& T^{3}_{}(S T^{2}_{})^{2}_{}, T^{3}_{} S T^{2}_{} S T^{4}_{}R, T^{4}_{} S T^{2}_{} S T^{3}_{}R; \nonumber \\
20C^{}_6 &:& S T R, T SR, S T^{4}_{}, T^{4}_{} S, T S T^{3}_{}, T^{2}_{} S T^{2}_{}, T^{2}_{} S T^{4}_{}R, T^{3}_{} S T, T^{3}_{} S T^{3}_{}R, T^{4}_{} S T^{2}_{}R,\nonumber \\
&& TST^3_{}S, T^2_{} S T^{3}_{} SR, T^{3}_{} S T^{2}_{} SR, S T^{2}_{} S T^{3}_{}R, S T^{3}_{} S T, S T^{3}_{} S T^{2}_{}R, (T^{2}_{} S)^{2}_{} T^{2}_{}, \nonumber \\
&&  T^{2}_{}(T^{2}_{} S)^{2}_{},(S T^{2}_{})^{2}_{} SR,(S T^{2}_{})^{2}_{} T^{2}_{}.
\label{eq:conclass}
\end{eqnarray}
The character table of $A^\prime_5$ has been shown in Table~\ref{table:character_table}, where $\phi\equiv (\sqrt{5}+1)/2$ has been defined. The irreducible representation matrices of three generators $S$, $T$ and $R$ are summarized as below
\begin{eqnarray}
{\bf 1} &: & \rho(S) = + 1 \; , \quad \rho(T) = + 1 \; , \quad \rho(R) = +1 \; , \nonumber \\
\widehat{\bf 2} &: & \rho(S)=\dfrac{\rm i}{\sqrt[4]{5}}\left(
\begin{matrix}
\sqrt{\phi} & \sqrt{\phi-1} \\
\sqrt{\phi-1} & -\sqrt{\phi} \\
\end{matrix}\right) \; , \quad
\rho(T)=\left(
\begin{matrix}
\omega^2_{} & 0 \\
0 & \omega^3_{} \\
\end{matrix}\right) \; , \quad
\rho(R)=- \mathbb{I}^{}_{2\times 2}\; , \nonumber \\
\widehat{\bf 2}^{\prime}_{} &: & \rho(S)=\dfrac{\rm i}{\sqrt[4]{5}}\left(
\begin{matrix}
\sqrt{\phi-1} & \sqrt{\phi} \\
\sqrt{\phi} & -\sqrt{\phi-1} \\
\end{matrix}\right) \; , \quad
\rho(T)=\left(
\begin{matrix}
\omega & 0 \\
0 & \omega^4_{} \\
\end{matrix}\right) \; , \quad
\rho(R)=- \mathbb{I}^{}_{2\times 2}\; , \nonumber  \\
{\bf 3} &:& \rho(S)=\dfrac{1}{\sqrt{5}}\left(\begin{matrix}
1 & -\sqrt{2} & -\sqrt{2} \\
-\sqrt{2} & -\phi & \phi-1 \\
-\sqrt{2} & \phi-1 & -\phi \\
\end{matrix}\right) \; , \quad \rho(T)=\left(\begin{matrix}
1 & 0 & 0 \\ 0 & \omega & 0 \\ 0 & 0 & \omega^4_{}
\end{matrix}\right) \; ,  \quad \rho(R)=+\mathbb{I}^{}_{3\times 3}\; , \nonumber  \\
{\bf 3}^{\prime}_{} &: & \rho(S)=\dfrac{1}{\sqrt{5}}\left(\begin{matrix}
-1 & \sqrt{2} & \sqrt{2} \\
\sqrt{2} & 1-\phi & \phi \\
\sqrt{2} & \phi & 1-\phi \\
\end{matrix}\right) \; , \quad \rho(T)=\left(\begin{matrix}
1 & 0 & 0 \\
0 & \omega^2_{} & 0 \\
0 & 0 & \omega^3_{} \\
\end{matrix}\right) \; , \quad \rho(R)=+\mathbb{I}^{}_{3\times 3}\; , \nonumber  \\
{\bf 4} &:& \rho(S) =  \dfrac{1}{\sqrt{5}}
\left(\begin{matrix}
1 & \phi-1 & \phi & -1 \\
\phi-1 & -1 & 1 & \phi \\
\phi & 1 & -1 & \phi-1 \\
-1 & \phi & \phi-1 & 1 \\
\end{matrix}\right) \; , \quad
\rho(T)=\left(\begin{matrix}
\omega & 0 & 0 & 0\\
0 & \omega^2_{} & 0 &  0\\
0 & 0 & \omega^3_{} &  0 \\
0 & 0 & 0 & \omega^4_{} \\
\end{matrix}\right)  \; , \quad \rho(R)=+\mathbb{I}^{}_{4\times 4}\; , \nonumber  \\
\widehat{\bf 4} &:& \rho(S) =  \dfrac{\rm i}{5^{\frac{3}{4}}_{}}
\left(\begin{matrix}
-\sqrt{2\phi+1} & \sqrt{3\phi} & \sqrt{3(\phi-1)} & \sqrt{2\phi-3} \\
\sqrt{3\phi} & \sqrt{2\phi-3} & \sqrt{2\phi+1} & \sqrt{3(\phi-1)} \\
\sqrt{3(\phi-1)} & \sqrt{2\phi+1} & -\sqrt{2\phi-3} & -\sqrt{3\phi} \\
\sqrt{2\phi-3} & \sqrt{3(\phi-1)} & -\sqrt{3\phi} & \sqrt{2\phi+1} \\
\end{matrix}\right) \; , \quad
\rho(T)=\left(\begin{matrix}
\omega & 0 & 0 & 0\\
0 & \omega^2_{} & 0 &  0\\
0 & 0 & \omega^3_{} &  0 \\
0 & 0 & 0 & \omega^4_{} \\
\end{matrix}\right)  \; , \nonumber  \\
&& \rho(R)=-\mathbb{I}^{}_{4\times 4}\; , \nonumber \\
{\bf 5} &:& \rho(S) =  \dfrac{1}{5}
\left(\begin{matrix}
-1 & \sqrt{6} & \sqrt{6} & \sqrt{6} & \sqrt{6} \\
\sqrt{6} & (\phi-1)^2_{} & -2\phi & 2(\phi-1) & \phi^2_{}\\
\sqrt{6} & -2\phi & \phi^2_{} & (\phi-1)^2_{} & 2(\phi-1) \\
\sqrt{6} & 2(\phi-1) & (\phi-1)^2_{} & \phi^2_{} & -2\phi \\
\sqrt{6} & \phi^2_{} & 2(\phi-1) & -2\phi & (\phi-1)^2_{}
\end{matrix}\right) \; , \quad
\rho(T)=\left(\begin{matrix}
1 & 0 & 0 & 0 & 0 \\
0 & \omega & 0 & 0 & 0\\
0 & 0 & \omega^2_{} & 0 &  0\\
0 & 0 & 0 & \omega^3_{} &  0 \\
0 & 0 & 0 & 0 & \omega^4_{} \\
\end{matrix}\right)  \; , \nonumber  \\
&& \rho(R)=+\mathbb{I}^{}_{5\times 5}\; ,  \nonumber  \\
\widehat{\bf 6} &: &  \rho(S)=\dfrac{-\rm i}{5^{\frac{3}{4}}_{}}
\left(\begin{matrix}
\sqrt{\phi} &-\sqrt{2(\phi-1)} &\sqrt{2\phi-3} &-\sqrt{2\phi+1} & \sqrt{2\phi}& \sqrt{\phi-1} \\
-\sqrt{2(\phi-1)} &-\sqrt{\phi-1} &\sqrt{2(\phi-1)} &\sqrt{2\phi} &\sqrt{\phi}  &\sqrt{2\phi}\\
\sqrt{2\phi-3} & \sqrt{2(\phi-1)} & \sqrt{\phi} & -\sqrt{\phi-1}& -\sqrt{2\phi} & \sqrt{2\phi+1} \\
-\sqrt{2\phi+1} & \sqrt{2\phi} & -\sqrt{\phi-1} & -\sqrt{\phi} & \sqrt{2(\phi-1)} & \sqrt{2\phi-3} \\
\sqrt{2\phi} & \sqrt{\phi} & -\sqrt{2\phi}& \sqrt{2(\phi-1)} & \sqrt{\phi-1} &  \sqrt{2(\phi-1)} \\
\sqrt{\phi-1} & \sqrt{2\phi} & \sqrt{2\phi+1} &  \sqrt{2\phi-3} &  \sqrt{2(\phi-1)} & -\sqrt{\phi} \\
\end{matrix} \right) \; ,  \nonumber   \\
& & \rho(T) =\left(\begin{matrix}
1 & 0 & 0 & 0 & 0 & 0\\
0 &  \omega & 0 & 0 & 0 & 0\\
0 &  0 & \omega^2_{} & 0  & 0 & 0\\
0 & 0 & 0 & \omega^3_{} & 0  & 0 \\
0 & 0 & 0 & 0 & \omega^4_{} & 0  \\
0 & 0 & 0 & 0 & 0 & 1\\
\end{matrix}\right) \; ,  \quad \rho(R) = -\mathbb{I}^{}_{6\times 6} \; ,
\label{eq:irrep}
\end{eqnarray}
where $\omega \equiv e^{2\pi {\rm i}/5}$ and $\mathbb{I}^{}_{n\times n}$ denotes the $n$-dimensional identity matrix. Notice that the above representation matrices are equivalent to those in Ref.~\cite{Hashimoto:2011tn} via unitary transformations.

\section{The Kronecker product rules of $A^{\prime}_{5}$}\label{app:B}
In this Appendix, we summarize the decomposition rules of the Kronecker products of any two nontrivial irreducible representations of $A^{\prime}_5$, namely, \\
\renewcommand\arraystretch{1.2}

\noindent
\begin{tabular}{L{6cm}L{6cm}L{6cm}}
	$\widehat{\bf 2} \otimes \widehat{\bf 2} = {\bf 1}^{}_{\rm a} \oplus {\bf 3}^{}_{\rm s}$ & $\widehat{\bf 2}^{\prime}_{} \otimes \widehat{\bf 2}^{\prime}_{} = {\bf 1}^{}_{\rm a} \oplus {\bf 3}^{\prime}_{\rm s}$
	& $\widehat{\bf 2} \otimes \widehat{\bf 2}^{\prime}_{} = {\bf 4}$ \\
	$\left \{
	\begin{array}{l}
	{\bf 1}^{}_{\rm a} : - \dfrac{\sqrt{2}}{2} \left(\alpha_1^{ } \beta_2^{ } - \alpha_2^{ } \beta_1^{ }\right) \\
	{\bf 3}^{}_{\rm s} : \dfrac{\sqrt{2}}{2} \left(
	\begin{array}{c}
	\alpha _1^{ } \beta
	_2^{ }+\alpha _2^{ } \beta _1^{ } \\
	-\sqrt{2} \alpha _2^{ } \beta _2^{ } \\
	\sqrt{2} \alpha _1^{ } \beta _1^{ } \\
	\end{array}
	\right)
	\end{array}  \right.$
	& $\left\{
	\begin{array}{l}
	{\bf 1}^{}_{\rm a} : \dfrac{\sqrt{2}}{2}\left(\alpha_{1}^{} \beta_{2}^{}-\alpha_{2}^{} \beta_{1}^{}\right) \\
	{\bf 3}^{\prime}_{\rm s} :  \dfrac{\sqrt{2}}{2}\left(\begin{array}{c}
	\alpha_{1}^{} \beta_{2}^{}+\alpha_{2}^{} \beta_{1}^{} \\
	-\sqrt{2} \alpha_{1}^{} \beta_{1}^{} \\
	\sqrt{2} \alpha_{2}^{} \beta_{2}^{}
	\end{array}\right)
	\end{array} \right.$
	& ${\bf 4} : \left(
	\begin{array}{c}
	\alpha _1^{ } \beta _2^{ } \\
	\alpha _2^{ } \beta _2^{ } \\
	-\alpha _1^{ } \beta _1^{ } \\
	\alpha _2^{ } \beta _1^{ } \\
	\end{array}
	\right)$
	\vspace{0cm}
\end{tabular}
\begin{tabular}{L{8.5cm}L{8.5cm}}
	$\widehat{\bf 2} \otimes {\bf 3} = \widehat{\bf 2} \oplus \widehat{\bf 4}$ & $\widehat{\bf 2}^{\prime}_{} \otimes {\bf 3}^{\prime}_{} = \widehat{\bf 2}^{\prime}_{} \oplus \widehat{\bf 4}$ \\
	$\left\{\begin{array}{l}
	\widehat{\bf 2} : \dfrac{\sqrt{3}}{3} \left(
	\begin{array}{c}
	-\alpha _1^{ }
	\beta _1^{ }+\sqrt{2} \alpha _2^{ } \beta _3^{ } \\
	\sqrt{2} \alpha _1^{ }
	\beta _2^{ }+\alpha _2^{ } \beta _1^{ } \\
	\end{array}
	\right) \\
	\widehat{\bf 4} : \dfrac{\sqrt{3}}{3} \left(
	\begin{array}{c}
	-\sqrt{3} \alpha _1^{ } \beta _3^{ } \\
	\sqrt{2} \alpha _1^{ } \beta _1^{ }+\alpha _2^{ }
	\beta _3^{ } \\
	-\alpha _1^{ }
	\beta _2^{ }+\sqrt{2} \alpha _2^{ } \beta _1^{ } \\
	-\sqrt{3} \alpha _2^{ } \beta _2^{ } \\
	\end{array}
	\right)
	\end{array}\right.$
	& $\left\{\begin{array}{l}
	\widehat{\bf 2}^{\prime}_{} : \dfrac{\sqrt{3}}{3}\left(\begin{array}{c}
	\alpha_{1}^{} \beta_{1}^{}+\sqrt{2} \alpha_{2}^{} \beta_{2}^{} \\
	\sqrt{2} \alpha_{1}^{} \beta_{3}^{}-\alpha_{2}^{} \beta_{1}^{}
	\end{array}\right) \\
	\widehat{\bf 4} : \dfrac{\sqrt{3}}{3} \left(\begin{array}{c}
	-\sqrt{2} \alpha_{1}^{} \beta_{1}^{}+\alpha_{2}^{} \beta_{2}^{} \\
	-\sqrt{3} \alpha_{2}^{} \beta_{3}^{} \\
	-\sqrt{3} \alpha_{1}^{} \beta_{2}^{} \\
	\alpha_{1}^{} \beta_{3}^{}+\sqrt{2} \alpha_{2}^{} \beta_{1}^{}
	\end{array}\right)
	\end{array}\right.$
	\vspace{0cm}
\end{tabular}\\
\begin{tabular}{L{8.5cm}L{8.5cm}}
	$\widehat{\bf 2} \otimes {\bf 3}^{\prime}_{} = \widehat{\bf 6}$ & $\widehat{\bf 2}^{\prime}_{} \otimes {\bf 3} = \widehat{\bf 6}$ \\
	$\widehat{\bf 6} : -\dfrac{\sqrt{2}}{2}\left(
	\begin{array}{c}
	\alpha _1^{ } \beta
	_3^{ }-\alpha _2^{ } \beta _2^{ } \\
	\sqrt{2} \alpha _2^{ } \beta _3^{ } \\
	-\sqrt{2} \alpha _1^{ } \beta _1^{ } \\
	\sqrt{2} \alpha _2^{ } \beta _1^{ } \\
	-\sqrt{2} \alpha _1^{ } \beta _2^{ } \\
	\alpha _1^{ } \beta
	_3^{ }+\alpha _2^{ } \beta _2^{ } \\
	\end{array}
	\right)$
	& $\widehat{\bf 6} : \left(\begin{array}{c}
	\alpha_{1}^{} \beta_{3}^{} \\
	-\alpha_{1}^{} \beta_{1}^{} \\
	-\alpha_{1}^{} \beta_{2}^{} \\
	-\alpha_{2}^{} \beta_{3}^{} \\
	\alpha_{2}^{} \beta_{1}^{} \\
	-\alpha_{2}^{} \beta_{2}^{}
	\end{array}\right) $
	\vspace{0cm}
\end{tabular}\\
\begin{tabular}{L{8.5cm}L{8.5cm}}
	$\widehat{\bf 2} \otimes {\bf 4} = \widehat{\bf 2}^{\prime}_{} \oplus \widehat{\bf 6}$ & $\widehat{\bf 2}^{\prime}_{} \otimes {\bf 4} =  \widehat{\bf 2} \oplus \widehat{\bf 6}$  \\
	$\left\{\begin{array}{l}
	\widehat{\bf 2}^{\prime}_{} : -\dfrac{\sqrt{2}}{2} \left(
	\begin{array}{c}
	\alpha _1^{ } \beta
	_4^{ }+\alpha _2^{ } \beta _3^{ } \\
	\alpha _1^{ } \beta
	_2^{ }-\alpha _2^{ } \beta _1^{ } \\
	\end{array}
	\right) \\
	\widehat{\bf 6}: \dfrac{\sqrt{2}}{2} \left(
	\begin{array}{c}
	-\sqrt{2} \alpha _1^{ } \beta _3^{ } \\
	-\alpha _1^{ } \beta
	_4^{ }+\alpha _2^{ } \beta _3^{ } \\
	\sqrt{2} \alpha _2^{ } \beta _4^{ } \\
	-\sqrt{2} \alpha _1^{ } \beta _1^{ } \\
	\alpha _1^{ } \beta
	_2^{ }+\alpha _2^{ } \beta _1^{ } \\
	\sqrt{2} \alpha _2^{ } \beta _2^{ } \\
	\end{array}
	\right)
	\end{array}\right.$
	& $\left\{\begin{array}{l}
	\widehat{\bf 2} : \dfrac{\sqrt{2}}{2}\left(\begin{array}{c}
	\alpha_{1}^{} \beta_{1}^{}+\alpha_{2}^{} \beta_{3}^{} \\
	\alpha_{1}^{} \beta_{2}^{}-\alpha_{2}^{} \beta_{4}^{}
	\end{array}\right) \\
	\widehat{\bf 6} : \dfrac{\sqrt{2}}{2} \left(\begin{array}{c}
	-\alpha_{1}^{} \beta_{4}^{}-\alpha_{2}^{} \beta_{1}^{} \\
	-\sqrt{2} \alpha_{2}^{} \beta_{2}^{} \\
	\alpha_{1}^{} \beta_{1}^{}-\alpha_{2}^{} \beta_{3}^{} \\
	-\alpha_{1}^{} \beta_{2}^{}-\alpha_{2}^{} \beta_{4}^{} \\
	\sqrt{2} \alpha_{1}^{} \beta_{3}^{} \\
	\alpha_{1}^{} \beta_{4}^{}-\alpha_{2}^{} \beta_{1}^{}
	\end{array}\right)
	\end{array}\right.$
\end{tabular}\\
\begin{tabular}{L{8.5cm}L{8.5cm}}
	$\widehat{\bf 2} \otimes \widehat{\bf 4} = {\bf 3} \oplus {\bf 5}$ & $\widehat{\bf 2}^{\prime}_{} \otimes \widehat{\bf 4} =  {\bf 3}^{\prime}_{} \oplus {\bf 5}$  \\
	$\left\{\begin{array}{l}
	{\bf 3}:-\dfrac{1}{2}\left(\begin{array}{c}
	\sqrt{2}\left(\alpha^{}_{1} \beta^{}_{3}-\alpha^{}_{2} \beta^{}_{2}\right) \\
	-\sqrt{3} \alpha^{}_{1} \beta^{}_{4}+\alpha^{}_{2} \beta^{}_{3} \\
	\alpha^{}_{1} \beta^{}_{2}+\sqrt{3} \alpha^{}_{2} \beta^{}_{1}
	\end{array}\right) \\
	{\bf 5}: \dfrac{1}{2}\left(\begin{array}{c}
	\sqrt{2}\left(\alpha^{}_{1} \beta^{}_{3}+\alpha^{}_{2} \beta^{}_{2}\right) \\
	\alpha^{}_{1} \beta^{}_{4}+\sqrt{3} \alpha^{}_{2} \beta^{}_{3} \\
	2 \alpha^{}_{2} \beta^{}_{4} \\
	-2 \alpha^{}_{1} \beta^{}_{1} \\
	-\sqrt{3} \alpha^{}_{1} \beta^{}_{2}+\alpha^{}_{2} \beta^{}_{1}
	\end{array}\right)
	\end{array}\right.$
	& $\left\{\begin{array}{l}
	{\bf 3}^{\prime}_{} : \dfrac{1}{2} \left(\begin{array}{c}
	\sqrt{2}\left(\alpha_{1}^{} \beta_{4}^{}+\alpha_{2}^{} \beta_{1}^{}\right) \\
	\alpha_{1}^{} \beta_{1}^{}+\sqrt{3} \alpha_{2}^{} \beta_{3}^{} \\
	-\sqrt{3} \alpha_{1}^{} \beta_{2}^{}-\alpha_{2}^{} \beta_{4}^{}
	\end{array}\right) \\
	{\bf 5} : \dfrac{1}{2} \left(\begin{array}{c}
	\sqrt{2}\left(\alpha_{1}^{} \beta_{4}^{}-\alpha_{2}^{} \beta_{1}^{}\right) \\
	-2 \alpha_{2}^{} \beta_{2}^{} \\
	-\sqrt{3} \alpha_{1}^{} \beta_{1}^{}+\alpha_{2}^{} \beta_{3}^{} \\
	\alpha_{1}^{} \beta_{2}^{}-\sqrt{3} \alpha_{2}^{} \beta_{4}^{} \\
	2 \alpha_{1}^{} \beta_{3}^{}
	\end{array}\right)
	\end{array}\right.$
	\vspace{0.6cm}
\end{tabular}
\begin{tabular}{L{8.5cm}L{8.5cm}}
	$\widehat{\bf 2} \otimes {\bf 5} = \widehat{\bf 4} \oplus \widehat{\bf 6}$ & $\widehat{\bf 2}^{\prime}_{} \otimes {\bf 5} = \widehat{\bf 4} \oplus \widehat{\bf 6}$ \\
\end{tabular}
\begin{tabular}{L{8.5cm}L{8.5cm}}
	$\widehat{\bf 2} \otimes \widehat{\bf 6} = {\bf 3}^{\prime}_{} \oplus {\bf 4} \oplus {\bf 5}$ & $\widehat{\bf 2}^{\prime}_{} \otimes \widehat{\bf 6} = {\bf 3} \oplus {\bf 4}  \oplus {\bf 5}$ \\
\end{tabular}\\
\begin{small}
	\begin{tabular}{L{5.7cm}L{5.7cm}L{5.7cm}}
		$\mathbf{3} \otimes \boldsymbol{3}=\mathbf{1}^{}_{\rm s} \oplus \boldsymbol{3}^{}_{\rm a} \oplus \boldsymbol{5}^{}_{\rm s}$ & ${\bf 3}^{\prime}_{} \otimes {\bf 3}^{\prime}_{} = {\bf 1}^{}_{\rm s} \oplus {\bf 3}^{\prime}_{\rm a} \oplus {\bf 5}^{}_{\rm s}$ & $ {\bf 3} \otimes {\bf 3}^{\prime}_{} = {\bf 4} \oplus {\bf 5}$ \\
\end{tabular}
	\begin{tabular}{L{8.5cm}L{8.5cm}}
		${\bf 3} \otimes {\bf 4} = {\bf 3}^{\prime}_{} \oplus {\bf 4} \oplus {\bf 5}$ & ${\bf 3}^{\prime}_{} \otimes {\bf 4} = {\bf 3} \oplus {\bf 4} \oplus {\bf 5}$ \\
\end{tabular}
	\begin{tabular}{L{8.5cm}L{8.5cm}}
		${\bf 3} \otimes \widehat{\bf 4} = \widehat{\bf 2} \oplus \widehat{\bf 4} \oplus \widehat{\bf 6}$ & ${\bf 3}^{\prime}_{} \otimes \widehat{\bf 4} = \widehat{\bf 2}^{\prime}_{} \oplus \widehat{\bf 4} \oplus \widehat{\bf 6}$ \\
\end{tabular}
	\renewcommand\arraystretch{1.25}
	\begin{tabular}{L{8.5cm}L{8.5cm}}
		${\bf 3} \otimes {\bf 5} = {\bf 3} \oplus {\bf 3}^{\prime}_{} \oplus {\bf 4} \oplus {\bf 5}$ & ${\bf 3}^{\prime}_{} \otimes {\bf 5} = {\bf 3} \oplus {\bf 3}^{\prime}_{} \oplus {\bf 4} \oplus {\bf 5}$ \\
\end{tabular}\\
	\begin{tabular}{L{8.5cm}L{8.5cm}}
		${\bf 3} \otimes \widehat{\bf 6} = \widehat{\bf 2}^{\prime}_{} \oplus \widehat{\bf 4} \oplus \widehat{\bf 6}^{}_{1} \oplus \widehat{\bf 6}^{}_{2}$ & ${\bf 3}^{\prime}_{} \otimes \widehat{\bf 6} = \widehat{\bf 2} \oplus \widehat{\bf 4} \oplus \widehat{\bf 6}^{}_1 \oplus \widehat{\bf 6}^{}_2 $ \\
\end{tabular}\\
	\renewcommand\arraystretch{1.25}
	\begin{tabular}{L{8.5cm}L{8.5cm}}
		${\bf 4} \otimes {\bf 4} = {\bf 1}^{}_{\rm s} \oplus {\bf 3}^{}_{\rm a} \oplus {\bf 3}^{\prime}_{\rm a} \oplus {\bf 4}^{}_{\rm s} \oplus {\bf 5}^{}_{\rm s}$ & $\widehat{\bf 4} \otimes \widehat{\bf 4} = {\bf 1}^{}_{\rm a} \oplus {\bf 3}^{}_{\rm s} \oplus {\bf 3}^{\prime}_{\rm s} \oplus {\bf 4}^{}_{\rm s} \oplus {\bf 5 }^{}_{\rm s} $ \\
\end{tabular}\\
	\begin{tabular}{L{17cm}}
		${\bf 4} \otimes \widehat{\bf 4} = \widehat{\bf 4} \oplus \widehat{\bf 6}^{}_1 \oplus \widehat{\bf 6}^{}_2$ \\
		$\left\{ \begin{array}{l}
		\widehat{\bf 4} :  \dfrac{\sqrt{5}}{5} \left(\begin{array}{c}
		-\sqrt{3} \alpha_{2}^{} \beta_{4}^{}-\alpha_{3}^{} \beta_{3}^{}+\alpha_{4}^{} \beta_{2}^{} \\
		-\alpha_{1}^{} \beta_{1}^{}-\alpha_{3}^{} \beta_{4}^{}+\sqrt{3} \alpha_{4}^{} \beta_{3}^{} \\
		-\sqrt{3} \alpha_{1}^{} \beta_{2}^{}+\alpha_{2}^{} \beta_{1}^{}-\alpha_{4}^{} \beta_{4}^{} \\
		\alpha_{1}^{} \beta_{3}^{}+\alpha_{2}^{} \beta_{2}^{}+\sqrt{3} \alpha_{3}^{} \beta_{1}^{}
		\end{array}\right) \\
		\widehat{\bf 6}^{}_1 :  \dfrac{1}{2}\left(\begin{array}{c}
		\alpha_{3}^{} \beta_{2}^{}-\sqrt{3} \alpha_{4}^{} \beta_{1}^{} \\
		-\sqrt{2} \left( \alpha_{3}^{} \beta_{3}^{}+ \alpha_{4}^{} \beta_{2}^{} \right) \\
		\sqrt{3} \alpha_{3}^{} \beta_{4}^{}+\alpha_{4}^{} \beta_{3}^{} \\
		\alpha_{1}^{} \beta_{2}^{}+\sqrt{3} \alpha_{2}^{} \beta_{1}^{} \\
		-\sqrt{2} \left( \alpha_{1}^{} \beta_{3}^{} - \alpha_{2}^{} \beta_{2}^{} \right) \\
		-\sqrt{3} \alpha_{1}^{} \beta_{4}^{}+\alpha_{2}^{} \beta_{3}^{}
		\end{array}\right) \\
		\widehat{\bf 6}^{}_2 :  \dfrac{\sqrt{2}}{4} \left(\begin{array}{c}
		\alpha_{1}^{} \beta_{4}^{}+\sqrt{3} \alpha_{2}^{} \beta_{3}^{}-2 \alpha_{4}^{} \beta_{1}^{} \\
		\sqrt{2} \left( \alpha_{2}^{} \beta_{4}^{}-\sqrt{3} \alpha_{3}^{} \beta_{3}^{} \right) \\
		2 \alpha_{1}^{} \beta_{1}^{}+\alpha_{3}^{} \beta_{4}^{}+\sqrt{3} \alpha_{4}^{} \beta_{3}^{} \\
		\sqrt{3} \alpha_{1}^{} \beta_{2}^{}+\alpha_{2}^{} \beta_{1}^{}-2 \alpha_{4}^{} \beta_{4}^{} \\
		\sqrt{2} \left( \sqrt{3} \alpha_{2}^{} \beta_{2}^{}- \alpha_{3}^{} \beta_{1}^{} \right)  \\
		-2 \alpha_{1}^{} \beta_{4}^{}-\sqrt{3} \alpha_{3}^{} \beta_{2}^{}-\alpha_{4}^{} \beta_{1}^{}
		\end{array}\right)
		\end{array}\right.$
		\vspace{0cm}
	\end{tabular}\\
	\begin{tabular}{L{17cm}}
		${\bf 4} \otimes {\bf 5} = {\bf 3} \oplus {\bf 3}^{\prime}_{} \oplus {\bf 4} \oplus {\bf 5}^{}_{1} \oplus {\bf 5}^{}_{2}$ \\
\end{tabular}\\
	\begin{tabular}{L{17cm}}
		$\widehat{\bf 4} \otimes {\bf 5} = \widehat{\bf 2} \oplus \widehat{\bf 2}^{\prime}_{} \oplus \widehat{\bf 4} \oplus \widehat{\bf 6}^{}_1 \oplus \widehat{\bf 6}^{}_2 $ \\
\end{tabular}\\
	\renewcommand\arraystretch{1.15}
	\begin{tabular}{L{17cm}}
		${\bf 4} \otimes \widehat{\bf 6} = \widehat{\bf 2} \oplus \widehat{\bf 2}^{\prime}_{} \oplus \widehat{\bf 4}^{}_1 \oplus \widehat{\bf 4}^{}_2 \oplus \widehat{\bf 6}^{}_1 \oplus \widehat{\bf 6}^{}_2 $ \\
\end{tabular}\\
	\begin{tabular}{L{17cm}}
		$\widehat{\bf 4} \otimes \widehat{\bf 6} = {\bf 3} \oplus {\bf 3}^{\prime}_{} \oplus {\bf 4}^{}_1 \oplus {\bf 4}^{}_2 \oplus {\bf 5}^{}_1 \oplus {\bf 5}^{}_{2} $ \\
\end{tabular}\\
	\renewcommand\arraystretch{1.23}
	\begin{tabular}{L{17cm}}
		${\bf 5} \otimes {\bf 5} = {\bf 1}^{}_{\rm s} \oplus {\bf 3}^{}_{\rm a} \oplus {\bf 3}^{\prime}_{\rm a} \oplus {\bf 4}^{}_{\rm s} \oplus {\bf 4}^{}_{\rm a} \oplus {\bf 5}^{}_{\rm s,1} \oplus {\bf 5}^{}_{\rm s,2} $\\
\end{tabular}\\
	\begin{tabular}{L{17cm}}
		${\bf 5} \otimes \widehat{\bf 6} = \widehat{\bf 2} \oplus \widehat{\bf 2}^{\prime}_{} \oplus \widehat{\bf 4}^{}_1 \oplus \widehat{\bf 4}^{}_2 \oplus \widehat{\bf 6}^{}_1 \oplus \widehat{\bf 6}^{}_2 \oplus \widehat{\bf 6}^{}_3$\\
\end{tabular}\\
	\begin{tabular}{L{17cm}}
		$\widehat{\bf 6} \otimes \widehat{\bf 6} = {\bf 1}^{}_{\rm a} \oplus {\bf 3}^{}_{\rm s,1} \oplus {\bf 3}^{}_{\rm s,2} \oplus {\bf 3}^{\prime}_{\rm s,1} \oplus {\bf 3}^{\prime}_{\rm s,2} \oplus {\bf 4}^{}_{\rm s} \oplus {\bf 4}^{}_{\rm a} \oplus {\bf 5}^{}_{\rm s} \oplus {\bf 5}^{}_{\rm a,1} \oplus {\bf 5}^{}_{\rm a,2}$\\
\end{tabular}%
\\
\end{small}

\renewcommand\arraystretch{1}

\end{document}